\documentclass[acmsmall]{acmart}

\def\BibTeX{{\rm B\kern-.05em{\sc i\kern-.025em b}\kern-.08emT\kern-.1667em\lower.7ex\hbox{E}\kern-.125emX}}

\setcopyright{acmcopyright}
\acmJournal{TOIT}
\acmVolume{X}
\acmNumber{X}
\acmArticle{X}
 \setcopyright{none}
 \renewcommand\footnotetextcopyrightpermission[1]{} % removes footnote with conference information in first column
\usepackage{algorithm}
\usepackage{algorithmic}
\usepackage{enumitem}

\begin{document}

%
% The "title" command has an optional parameter, allowing the author to define a "short title" to be used in page headers.
\title{Layer-based Composite Reputation Bootstrapping}

%
% The "author" command and its associated commands are used to define the authors and their affiliations.
% Of note is the shared affiliation of the first two authors, and the "authornote" and "authornotemark" commands
% used to denote shared contribution to the research.

\author{Sajib Mistry}
\affiliation{%
  \institution{Discipline of Computing, School of EECMS, Curtin University}
  %\streetaddress{J12/1 Cleveland St}
  %\city{Sydney}
  \state{WA}
  \postcode{6102}
  \country{Australia}}
\email{sajib.mistry@curtin.edu.au}

\author{Lie Qu}
%\orcid{1234-5678-9012-3456}
\affiliation{%
  \institution{Alibaba Group}
  %\streetaddress{J12/1 Cleveland St}
  %\city{Sydney}
  %\state{NSW}
  %\postcode{2008}
  \country{China}}
\email{E-mail: lie.qu@hotmail.com}

\author{Athman Bouguettaya}
\affiliation{%
  \institution{School of Computer Science, University of Sydney}
  %\streetaddress{J12/1 Cleveland St}
  %\city{Sydney}
  \state{NSW}
  \postcode{2008}
  \country{Australia}}
\email{athman.bouguettaya@sydney.edu.au}

%
% By default, the full list of authors will be used in the page headers. Often, this list is too long, and will overlap
% other information printed in the page headers. This command allows the author to define a more concise list
% of authors' names for this purpose.
\renewcommand\shortauthors{S. Mistry. et al}

%
% The abstract is a short summary of the work to be presented in the article.
\begin{abstract}
We propose a novel generic reputation bootstrapping framework for composite services. Multiple reputation-related indicators are considered in a layer-based framework to implicitly reflect the reputation of the component services. The importance of an indicator on the future performance of a component service is learned using a modified Random Forest algorithm. We propose a topology-aware Forest Deep Neural Network (fDNN) to find the correlations between the reputation of a composite service and reputation indicators of component services. The trained fDNN model predicts the reputation of a new composite service with the confidence value. Experimental results with real-world dataset prove the efficiency of the proposed approach.
\end{abstract}

%\begin{CCSXML}
%<ccs2012>
%<concept>
%<concept_id>10002951.10003260.10003304.10003306</concept_id>
%<concept_desc>Information systems~Web services</concept_desc>
%<concept_significance>300</concept_significance>
%</concept>
%<concept>
%<concept_id>10002951.10003227.10003233.10003449</concept_id>
%<concept_desc>Information systems~Reputation systems</concept_desc>
%<concept_significance>300</concept_significance>
%</concept>
%<concept>
%<concept_id>10002978.10002986.10002987</concept_id>
%<concept_desc>Security and privacy~Trust frameworks</concept_desc>
%<concept_significance>300</concept_significance>
%</concept>
%</ccs2012>
%\end{CCSXML}
%\ccsdesc[300]{Information systems~Web services}
%\ccsdesc[300]{Information systems~Reputation systems}
%\ccsdesc[300]{Security and privacy~Trust frameworks}

%
% Keywords. The author(s) should pick words that accurately describe the work being
% presented. Separate the keywords with commas.

%
% A "teaser" image appears between the author and affiliation information and the body
% of the document, and typically spans the page.
%%  \includegraphics[width=\textwidth]{sampleteaser}
%%  \caption{Seattle Mariners at Spring Training, 2010.}
%%  \Description{Enjoying the baseball game from the third-base seats. Ichiro Suzuki preparing to bat.}

%
% This command processes the author and affiliation and title information and builds
% the first part of the formatted document.
\maketitle

\section{Introduction}
The \textit{on-demand} service composition is a natural occurrence in a highly dynamic service market, e.g., Cloud, Micro-services and Business Process Management \cite{4279708}. The dynamic service environments are characterized by changing conditions in the presence of massive growth of new services, users' long-term QoS requirements, and QoS fluctuations in existing services \cite{erl2004}. In the dynamic service composition environment, atomic services collaboratively create on-demand value-added services following a composition topology. On-demand compositions are usually \textit{context-aware} and adapt to the changes in user's requirements \cite{7401117}. If a single service does not have the functionality to meet a consumer's requirements, related services are composed on the fly \cite{xliu2013}. It is challenging to access the \textit{initial reputation values of on-demand composite services} as they usually have little or no direct consumer feedback. This is referred to as the \textit{reputation bootstrapping} problem \cite{malik2009reputation}. 

\textit{Our objective is to build a reputation bootstrapping framework for a new on-demand composite service.} We assume that the component or atomic services are already selected in a given composition topology. For example, an ``airline search service" and a ``ticket price prediction service'' can be composed in a ``travel booking service''. If ``Kayak\footnote{\url{https://www.kayak.com/affiliate/widget-v2.js}}'' and ``AirHint\footnote{\url{https://www.airhint.com}}'' are two implementations of the ``search service" and a ``price prediction service'' respectively, our target is to bootstrap the reputation of the ``Airline booking service" which is composed of the atomic services\footnote{The terms atomic and component service are applied interchangeably throughout the paper}, i.e.,``Kayak'' and ``AirHint'' on the fly.

%The ubiquity of applications of Service-Oriented Architecture (SOA) \cite{erl2004} has elicited the emergence of various types of services to satisfy consumers' requirements. Due to the diversity of services, selecting the best-performed service before actual consumption is usually a big concern for consumers. In this regard, \textit{service performance} is an indication of reputation which may be deduced from previous interaction-based experience. Unfortunately, this information is usually unavailable for new service entrants making it a challenge to accurately compute the initial reputation value. This is referred to as the \textit{reputation bootstrapping} problem \cite{malik2009reputation}. 
 %In this regard, \textit{reputation} is an effective way to determine the performance quality of a service based on prior performance experiences (or records). However, performance experiences may not be always available when a new service emerges in a service market. Consequently, its reputation cannot be assessed, and thus trust establishment between consumers and the new service becomes challenging. \textit{Reputation bootstrapping} is a key enabler to assign appropriate initial reputations for \textit{new services} \cite{malik2009reputation}.

To the best of our knowledge, existing work on reputation bootstrapping mainly focus on \textit{atomic services} \cite{malik2009reputation,wu2015neural,nguyen2012bootstrapping,DBLP:conf/wise/SkopikSD09,DBLP:journals/sigecom/MaximilienS02,DBLP:conf/atal/HuynhJS06,tibermacine2015regression,8694843,8108459,2792979,3369390}. Atomic services are standalone services which are not dependent on other services \cite{malik2009reputation}. Empirical assumptions are that \textit{implicit performance} factors ought to be employed to determine the reputation of a new service.
%These approaches are typically based on particular empirical assumptions in which the reputation of a new service can be extracted from some \textit{implicit performance indicators}, which are usually the inherent characteristics of services. 
%For example, the approach proposed in \cite{nguyen2012bootstrapping} discovers that a new service provided by a \textit{reputable} provider tends to offer \textit{good} performance. The reputation of the provider can be applied to estimate the future performance of a new service. 
There have been suggestions that a new service tends to have a good performance if the provider is \textit{reputable} \cite{nguyen2012bootstrapping}. In this case, the reputation of the provider can be used as a good estimator for the future performance of a new service. The reputation of a new service has a strong correlation to the reputation of similar services in specific contexts \cite{DBLP:conf/wise/SkopikSD09}. In this paper, the term ``service similarity'' indicates the similarity of both functional and non-functional, i.e., QoS aspects between two services \cite{malik2009reputation}.   
%The approach proposed in \cite{DBLP:conf/wise/SkopikSD09} indicates that the reputation of a new service has a strong correlation to the reputation of similar services in specific contexts. 
By calculating \textit{service similarity} among services, a new service's reputation can be predicted from a cluster of similar services. However, \textit{these specific assumptions may only hold for particular cases but are not always true in a generic way}. 

A service usually has multiple performances or reputation indicators (e.g., its provider, community, or similar services), each of which can \textit{relatively} reflect the future performance of the provider to some extent \cite{qu2017confidence}. The \textit{correlation} between the services reputation and each indicator is usually \textit{dynamic}. For example, if a super shop retailer introduces new gaming services (gaming PCs, servers, etc.) to its consumers without previous experience in electronics, the reputation of the \textit{provider} for selling everyday objects should not be a major performance indicator to bootstrap the reputation of the new gaming services. The reputation of \textit{community} (i.e., the reputation of other super shops in consumer electronics) may have a higher correlation to the bootstrapped reputation of the new gaming service. The reputation bootstrapping should analyze the relative importance \textit{quantitatively} among different indicators, i.e., which indicators can more effectively reflect new services' future reputations. The effectiveness of reputation indicators can help consumers to evaluate the \textit{reliability} of reputation bootstrapping under various circumstances. In our previous work \cite{qu2017confidence}, we have proposed a layer-based framework to determine how much the reputation indicators can influence the reputation of an \textit{atomic} service.

\textit{Our objective is to bootstrap the reputation of a composite service}. %The service composition provides an elegant means to build a value-added service invoking other component services \cite{4279708}. For example, a ``voice-enabled navigation" service can be composed of a ``map" service, a ``traffic analysis" service and a ``voice assistant" service. 
One simple approach is to represent the minimum reputation of component services as the reputation of a composite service. This approach may not reflect the importance of each component.  For example, the ``weakest link'' may be related to the least significant component in the composed service.  Therefore, assigning the overall reputation in this case to the least significant component would not be accurate.  In this respect, each component service may have different levels of contributions towards the overall reputation of the composite service \cite {nepal2009reputation, 9328204}. A component service may contribute less in comparison to other component services. For example, a ``voice assistant service'' providing hands-free route selection may have less impact on the overall perception of a ``map service'', in comparison with the component ``traffic analysis service''. If the reputation of the ``voice assistant service'' is the minimum, it would be inaccurate to assign the reputation of the ``map service'' only considering the ``voice assistant service'' as the effect of ``traffic analysis service'' are not considered.

%However, this approach may may not be applicable universally as each component service may have different levels of contributions towards the overall reputation of the composite service \cite{nepal2009reputation}. A component service may contribute less in comparison to other component services. For example, a ``voice assistant service'' providing hands-free route selection may have less impact on the overall perception of a ``map service'', in comparison with the component ``traffic analysis service''. If the reputation of the “voice assistant service” is the minimum, it would be inaccurate to assign the reputation of the ``map service'' only considering the “voice assistant service” as the effects of ``traffic analysis service'' are not considered.     

The reputation bootstrapping for composite services is more \textit{challenging} than that for atomic services. 
%One of the key issues is that the correlation between the performance of a composite service and that of its corresponding component services is usually \textit{dynamic} and may \textit{change} according to the composition topology \cite{xliu2013}. 
A key issue is the \textit{dynamic correlation} between the performance of a composite service and that of its components. This is dependent on the composition topology \cite{xliu2013}. \textit{In this paper, we only consider acyclic topologies, i.e., the pointed or invoked service does not link back to the invoking service in the composition.} This assumption is without loss of generality as cyclic graphs can always be transformed into acyclic graphs \cite{ul2010aggregation,neiroukh2008transforming}. Note, any graph-based cycle detection algorithm and directed acyclic graphs (DAGs) software modeling, e.g., MIM, and Tetrad could be applied to determine cycles in a composition topology \cite{haughton2006review}. Due to compatibility issues, the composition of reputable atomic or component services may not always induce a well-performed composite service. Another challenge is that the \textit{relative influence} among multiple performance indicators of component services is {context-aware} \cite{yamato2006context}. A component service may have different levels of reputation influence in a composition. The same component service invoked in different compositions may perform inconsistently in different contexts. For example, the ``Google Map service'' provides a stripped down performance in the user experience when invoked by the voice command service Siri in Apple carplay. However, the performance increases in Android Auto where Google Map services are seamlessly composed with ``Ok Google'' voice services (Google map user ratings for Apple carplay and Android Auto is 4 and 5 respectively) \cite{Mashable}. 

The performance of a composite service should be estimated by not only the individual performance of its component services but also their compatibility in a service composition. We term such a phenomenon as ``\textit{performance interdependence}'', i.e., the performance of a composite service may be influenced by the topology of the composition. The performance interdependence may be caused by different factors. For example, the component services may not be \textit{properly invoked} as their specifications. There may exist functional \textit{compatibility} issues among component services, e.g., two component services follow different data transfer protocols, and thus cannot communicate smoothly. The performance correlation between component services and the corresponding composite service should be carefully studied when bootstrapping the reputations of composite services. Note that, a component service in a composition can be composed of other atomic services. In such case, the whole reputation bootstrapping of composite services may be \textit{recursive}, i.e., we need to bootstrap the reputation of component services at first. In this paper, \textit{we only consider non-recursive reputation bootstrapping of a composition service which is composed of atomic services}. The reputation-related factors of each component services are determined using the layered-based approach of bootstrapping reputation for a single service \cite{nguyen2012bootstrapping,qu2017confidence}. The required information, i.e., reputation-related factors for component services are assumed to have been already generated. 

We leverage the \textit{reputation indicators} and \textit{topology-based interactions} among component services to bootstrap reputation of a composite services. The term ``topology-aware'' in a reputation bootstrapping model implies that the composite reputation classification model is directed by the composition shape (i.e., topology) and considers interactions among the component services and complex correlations of the reputation indicators among the component services. In contrast, Topology Free Reputation Bootstrapping (TFRB) is an initial approach where we consider only the component services but discard the composition topology. This topology-aware approach is a \textit{passive} way of reputation bootstrapping, i.e., bootstrapping based on existing information. The reliability of the bootstrapped reputation depends on the \textit{quality} of the existing information. For example, if the provider of a newly developed service has no historical records, and the provider does not belong to a community, i.e., similar providers do not exist, the performance indicator ``\textit{provider}'' is invalid for bootstrapping the service's reputation. In such cases, an intuitive idea is to apply a trial period to demonstrate the performance of the service \cite{jiao2011framework,malik2009reputation}. The trial-based approach can be considered as an \textit{active} way of reputation bootstrapping, i.e., let service positively prove its actual performance. Note that, the trial-based approaches are not an alternative to the topology-based approach, rather they are \textit{complementary} in nature. Some service providers may not offer free trials which incur additional costs. It should also be guaranteed that the services attending in the trials are not specifically designed to perform better. It would also take a large amount of time to try every component service in a composition. Due to the consideration of efficiency, the reputations of most of the new services should be predicted using the indicator and topology-based approach. The trial-based approach should be considered when the \textit{confidence} of bootstrapped reputations are low, i.e., the prediction is unreliable. The confidence of every bootstrapped reputation should be quantitatively modeled before conducting trials. The lower confidence means the leveraged performance indicators or the composition topology are ineffective for reputation bootstrapping and the service requires to attend a performance trial.

We propose a \textit{generic bootstrapping framework} to determine the reputation of the on-demand service composition in a highly dynamic service market, e.g., Cloud, Micro-services and Business Process Management \cite{4279708}. The dynamic service environments are characterized by changing conditions in the explosive growth of new services, users' long-term QoS requirements, and QoS fluctuations in existing services \cite{xliu2013}. In the dynamic service composition environment, atomic services collaboratively create on-demand value-added services following a composition topology on the fly. It is challenging to determine the \textit{reputation of on-demand composite services} as they usually have little or no direct consumer feedback. The existing approaches of reputation bootstrapping for atomic services are not directly applicable for on-demand compositions as they do not consider the topology-based interactions and the performance interdependence among the atomic services \cite{malik2009reputation,wu2015neural,nguyen2012bootstrapping,DBLP:conf/wise/SkopikSD09,DBLP:journals/sigecom/MaximilienS02,DBLP:conf/atal/HuynhJS06}. One of the key contributions of the research is that the proposed framework does not make any concrete \textit{empirical} assumption for reputation bootstrapping (e.g., only use providers' reputations as their services' reputation). Instead, we apply state-of-art \textit{data-driven machine learning} approaches to make it self-adapt to different concrete circumstances by leveraging the reputation indicators and the composition topology to determine the reputation of a composite service. We also propose a model to evaluate the \textit{confidence} of the bootstrapped reputation of the composite service. Our contributions in this paper are summarized as follows:

% \begin{itemize}[topsep=0pt,parsep=0pt,partopsep=0pt,leftmargin=10pt,labelwidth=6pt,labelsep=4pt]
\begin{itemize}
[topsep=0pt,parsep=0pt,partopsep=0pt,leftmargin=10pt,labelwidth=6pt,labelsep=4pt]
    \item We propose a novel \textit{generic layer-based} and \textit{composition topology-aware} reputation bootstrapping framework for composite services. %The proposed framework includes a \textit{layer-based} reputation bootstrapping model for atomic services and a topology-aware reputation bootstrapping model for composite services.
    \item  The proposed framework does not rely on \textit{empirical} assumptions. A data-driven approach using random forest \cite{liaw2002classification} is proposed to determine the importance of reputation-related indicators quantitatively. %The proposed \textit{layer-based} reputation bootstrapping model considers multiple reputation-related indicators which may implicitly reflect a newly developed atomic services' future performance. It does not rely on \textit{empirical} assumptions. A data-driven approach using random forest \cite{liaw2002classification} is proposed to determine the importance of reputation-related indicators quantitatively. 
   \item A new approach is proposed to learn the reputation influence among component services from historical records using the forest deep neural network (fDNN) \cite{kong2018deep}. %The fDNNs are chained and trained following the composition topology to capture the hidden correlations among the reputation indicators of the component services.
   %\item We quantitatively define \textit{confidence} which describes the reliability of new services' bootstrapped reputations. The notion \textit{confidence} helps to identify the services whose bootstrapped reputation cannot be effectively predicted. Such services are required to attend a trial to demonstrate their actual performance.
    \item We conduct experiments based on a \textit{real-world} dataset from GitHub to evaluate the accuracy and efficiency of the proposed approach.
\end{itemize}

\subsection{Motivating Scenario}
\begin{figure}[t!]
\vspace{-.5cm}
\begin{minipage}{0.55\textwidth}
%\centering
%\begin{table}
    
   \resizebox{1\textwidth}{!}{  
    		\begin{tabular}{|p{3.5cm}|p{6cm}|}
    		%\begin{tabular}{|l|l|}
			\hline
			\textbf{Abstract Services} & \textbf{Concrete Services in GitHub} \\ 
			\hline
			Multi Room Music System (MRMS) & soulshake/fonos, tomtaylor/multiroom-audio  \\ 
			\hline
			Multi Remote Controller (MRC) & xkonni/raspberry-remote, kdvolder/omx-remote \\ \hline
			Web Client for MRMS (WC) & pimusicbox/mopidy-musicbox-webclient, CAtOSe/mopidy-minimal-webclient \\
			\hline
			Online Integration System platform for MRMS (OIS) & opsxcq/docker-mopidy, davidarkemp/mopidy-mixcloud \\
            
            \hline
			YouTube based OIS & mopidy/mopidy-youtube, SKrotkih/YTLiveStreaming \\
			\hline
			Spotify based OIS & mopidy/mopidy-spotify, spotify/web-api \\ 
			\hline
			iTunes based OIS & agg23/Mopidy-AppleMusic,	PouleR/apple-music-api  \\
			\hline
			Music Server System (MSS) & mopidy/mopidy,	phanan/koel  \\ 
			\hline
			Concurrent Networking Library (CNL) & jodal/pykka, soravux/scoop \\ 
			\hline
			Security Validation (SV)  & OWASP/ASVS,	omg-dds/dds-security \\
			\hline
		\end{tabular}
		
		}
		\caption{Abstract Services and Concrete Services in GitHub}
    \label{tb1}
%\end{table}
\end{minipage}
\hspace{0.4cm}
\begin{minipage}{0.40\textwidth}
%\begin{figure}[b]
    \centering
    \vspace{2.4cm}
    \includegraphics[width=1\textwidth,height=.7\textwidth]{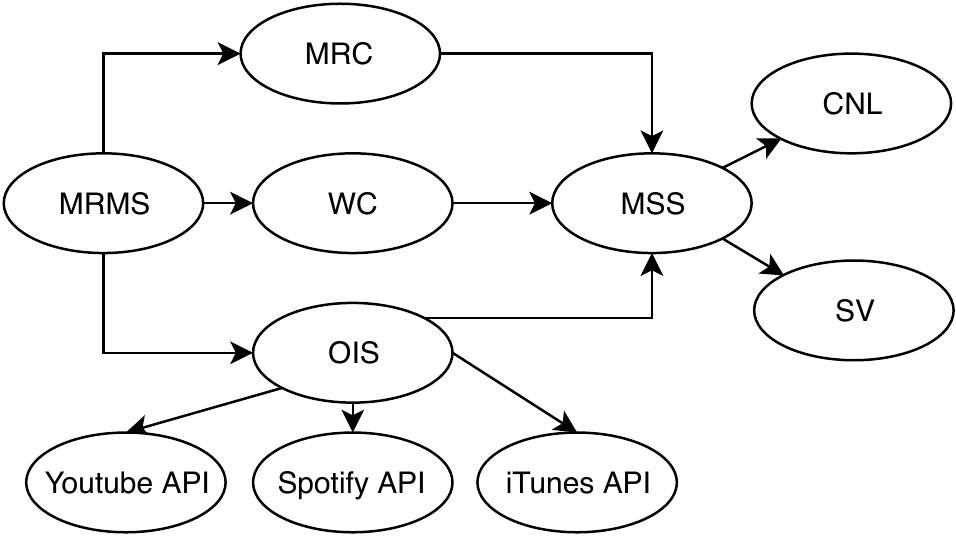}
	\caption{The composition topology for MRMS}
	\label{fig:mot}
%\end{figure}
\end{minipage}
\vspace{-.3in}
\end{figure}

The web-based hosting service GitHub\footnote{\url{https://github.com}} is an excellent opensource platform to develop new software services composing existing project repositories. In GitHub, a repository can comprise one or more repositories as submodules and create a dependency graph. However, we only consider repositories with acyclic dependency graph. Let us assume, a software company wants to develop a Multi-Room Music System (MRMS) in GitHub. The objective of the MRMS is to play music on sound systems (located in different rooms) from local devices, as well as online music streaming services such as Spotify, YouTube etc. Examples of MRMS repositories in GitHub include `soulshake/fonos'\footnote{\url{https://github.com/soulshake/fonos}} and `tomtaylor/multiroom-audio'\footnote{\url{https://github.com/tomtaylor/multiroom-audio}}. The MRMS consists of several \textit{abstract} modules or software services. Each of the abstract services (specifications of functionality and QoS) is implemented via a concrete software service which hosted in GitHub. Figure \ref{tb1} lists the required abstract services and corresponding concrete services. For example, `mopidy/mopidy'\footnote{\label{mopidy}\url{https://github.com/mopidy/mopidy}} and `phanan/koel'\footnote{\url{https://github.com/phanan/koel}} are the concrete implementations of the Music Server System (MSS).

% % \begin{table}[t!]
% %     \caption{Abstract Services and Concrete Services in GitHub for MRMS}
% %     \resizebox{.55\textwidth}{!}{  
% %     		\begin{tabular}{|p{3.5cm}|p{6cm}|}
% % 			\hline
% % 			\textbf{Abstract Services} & \textbf{Concrete Services in GitHub} \\ 
% % 			\hline
% % 			Multi Room Music System (MRMS) & soulshake/fonos, tomtaylor/multiroom-audio  \\ 
% % 			\hline
% % 			Multi Remote Controller (MRC) & xkonni/raspberry-remote, kdvolder/omx-remote \\ \hline
% % 			Web Client for MRMS (WC) & pimusicbox/mopidy-musicbox-webclient, CAtOSe/mopidy-minimal-webclient \\
% % 			\hline
% % 			Online Integration System platform for MRMS (OIS) & opsxcq/docker-mopidy, davidarkemp/mopidy-mixcloud \\
            
% %             \hline
% % 			YouTube based OIS & mopidy/mopidy-youtube, SKrotkih/YTLiveStreaming \\
% % 			\hline
% % 			Spotify based OIS & mopidy/mopidy-spotify, spotify/web-api \\ 
% % 			\hline
% % 			iTunes based OIS & agg23/Mopidy-AppleMusic,	PouleR/apple-music-api  \\
% % 			\hline
% % 			Music Server System (MSS) & mopidy/mopidy,	phanan/koel  \\ 
% % 			\hline
% % 			Concurrent Networking Library (CNL) & jodal/pykka, soravux/scoop \\ 
% % 			\hline
% % 			Security Validation (SV)  & OWASP/ASVS,	omg-dds/dds-security \\
% % 			\hline
% % 		\end{tabular}
% % 		}
% %     \label{tb1}
% % \end{table}

% \begin{figure}[b]
%     \centering
%     \includegraphics[width=.55\textwidth,height=.3\textwidth]{figures/motivation.pdf}
% 	\caption{The composition topology for MRMS}
% 	\label{fig:mot}
% \end{figure}

If a service invokes another service, we use a \textit{directed} edge to represent the direction of service invocations. The acyclic composition topology of the MRMS is described in Figure \ref{fig:mot}. The MRMS requires the Multi-Remote Controller (MRC) to synchronize the Raspberry Pi music devices in different rooms, and the Web Client (WC) to build the user interface and an Online Integration System (OIS) to connect to the internet and stream music from different platforms such as Spotify and iTunes.  All these services should be built upon an extensible Music Server System (MSS) that supports to add new music sources easily. The MSS is comprised of Concurrent Network Library (CNL) to easily manage the concurrent applications and Security Validation (SV) services to ensure the integrity among different concurrent applications. The OIS requires application programming interfaces (API) of different streaming platforms (i.e., YouTube, Spotify, and iTunes) to connect to those services and play music. Note that the pointed or invoked service does not link back to the invoking service in Figure 2.

We can \textit{create} different concrete compositions by selecting the corresponding concrete services for each of the abstract services from (Figure \ref{tb1}) in the composition topology (Figure \ref{fig:mot}). Let us assume a concrete composition for MRMS is created which is  \textit{A} = \{xkonni/raspberry-remote, pimusicbox/mopidy-musicbox-webclient, opsxcq/docker-mopidy,  mopidy/mopidy-youtube, mopidy/mopidy-spotify, mopidy/mopidy, agg23/Mopidy-AppleMusic, jodal/pykka, OWASP/ASVS\}. Another example of the composition is \textit{B} = \{kdvolder/omx-remote,  CAtOSe/mopidy-minimal-webclient, soravux/scoop, davidarkemp/mopidy-mixcloud,  SKrotkih/YTLiveStreaming, spotify/web-api, PouleR/apple-music-api, phanan/koel,  omg-dds/dds-security\}. The objective is to \textit{estimate the reputation} of the compositions. Note that both \textit{A} and \textit{B} are \textit{on-demand} composite services, i.e., newly created. Estimating a concrete on-demand MRMS reputation is quite \textit{challenging} because the \textit{direct reputation} records of concrete component services may be missing if they are newly developed in GitHub.

The reputation of a GitHub repository can be bootstrapped from the reputations of their contributors, owners or communities or similar services. For example, the repository `mopidy/mopidy\textsuperscript{\ref{mopidy}}' is in operation for 1 year and has 5670 stars or ratings from users which are accumulated, 237 watchers, 261 dependent repositories, 13 dependencies, 106 contributors, the community or owner has 22 other repositories with average 1000 ratings or stars and 120 followers. The GitHub also provides \textit{insights} about the commits, code frequency, and forks of these repositories.  %Similarly, the repository `xkonni/raspberry-remote\footnote{\url{https://github.com/xkonni/raspberry-remote}}' is in operation for 6 years and has 88 stars or ratings from users, 20 watchers, 0 dependent repository, 0 dependencies, 9 contributors, the community or owner has 31 other repositories with average 40 ratings or stars and 31 followers. The Github also provides \textit{insights} about other reputation indicators such as the commits, code frequency and forks of these repositories. 

Although GitHub provides different reputation indicators, it is challenging to  \textit{quantitatively aggregating} these indicators to reflect a component's actual reputation. The aggregation strategy for one type of repository may not be applicable to different types of repositories. As the concrete component services are individually developed by different contributors, their compatibility may affect their performance. The concrete component services may be developed by different programming languages or may follow different programming specifications (e.g., different security standards). This may lead to many unpredictable compatibility issues. It would be difficult to predict the composite reputation without considering the composition topology. Hence, the reputation bootstrapping should consider all these factors and learn an effective model according to the particular situation of service compositions. 
\section{The Layer-based Reputation Bootstrapping for Atomic Services}
\subsection{The Layer-based Reputation Transfer}

% \begin{figure}
% 	\centering
% 	\begin{center}
% 		%\scalebox{.8}{\includegraphics{figures/reputation-transfer.pdf}}
% 		\includegraphics[width=.75\textwidth]{figures/reputation-transfer.pdf}
% 	\end{center}
% 	%\vspace{-.1in}
% 	\caption{The layer-based reputation transfer}
% 	%\vspace{0in}
% 	\label{Fig3}
% 	%\vspace{-.3in}
% \end{figure}

Existing approaches identify some of the key reputation indicators as follows:
\begin{itemize}
[topsep=0pt,parsep=0pt,partopsep=0pt,leftmargin=10pt,labelwidth=6pt,labelsep=4pt]
\item \textit{Provider}: The reputation of the service provider could be transferred as the bootstrapped reputation of a new service \cite{DBLP:conf/wise/SkopikSD09}. For example, as Samsung has a higher reputation in mobile screen manufacturing, higher performance is expected from a new foldable screen developed by Samsung. In the context of GitHub, a \textit{contributor} of a repository may be regarded as a provider.
\item \textit{Community}: The reputation of the new service could be highly correlated to the reputation of the community \cite{DBLP:journals/sigecom/MaximilienS02}. A service belonging to a reputable community has a higher probability of providing good services \cite{DBLP:conf/atal/HuynhJS06}. A community could be the \textit{owner} of a repository in GitHub.

\item \textit{Similar Services}: If most of the similar services in the market could be clustered based on performances, there is a higher probability that the performance of the new service will be similar to the performance of the corresponding cluster of similar services \cite{yahyaoui2011bootstrapping}. For example, if budget airlines have limited access to WIFI services, a newly created budget airline has a higher probability to provide similar internet experience to its consumers. 
\end{itemize} 

\begin{figure}[t!]
%\vspace{-.2in}
    \begin{minipage}{0.58\textwidth}
    	\centering
		\includegraphics[width=1\textwidth]{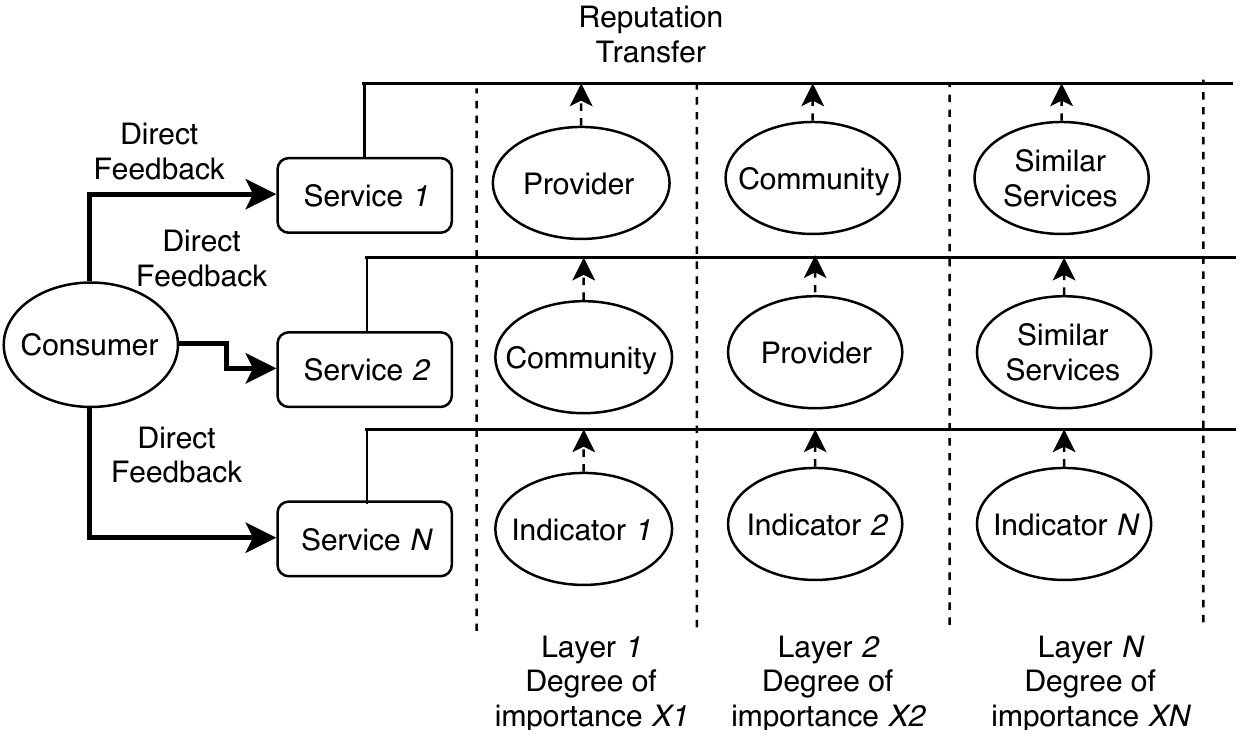}
	    \caption{The layer-based reputation transfer}
	\label{Fig3}
    \end{minipage}
    \begin{minipage}{0.40\textwidth}
   
	\centering
	\includegraphics[width=.9\textwidth]{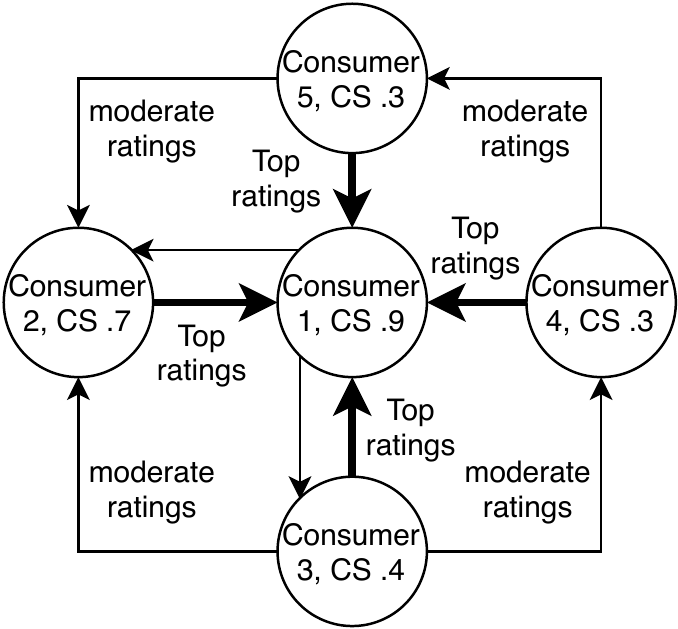}
	\caption{Credibility-based peer to peer direct assessment}
	\label{Fig4}
    \end{minipage}
    \vspace{-.15in}
\end{figure}

The \textit{insight} or meta-information of service is also a reputation indicator. In GitHub, examples of insight are watchers, dependencies, dependents, forks, network, commit timelines, bug fixing, documentation, Wiki and so on. Let us assume that a set of \textit{N} reputation indicators for service \textit{X} is given as \textit{RI}=\{Indicator 1, Indicator 2, $\dots$ , Indicator N\}. The objective is to find the \textit{degree of importance} of each reputation indicator in the bootstrapping process. %We propose a \textit{layer-based framework} to quantitatively model the importance of \textit{RI}. 
Figure \ref{Fig3} describes the proposed framework of reputation transfer. We consider two types of transfer: a) \textit{direct feedback from consumers}, and  b) \textit{indirect reputation transfer from indicators}. The direct feedback from the consumers or ratings could be interpreted as the reputation of the service. It should be ensured the ratings are genuine and are not deliberately tempered. We consider these direct ratings as the \textit{ground truth} for bootstrapping reputation of new services. 

The new services usually do not have enough direct feedback. The layer-based framework (Figure \ref{Fig3}) \textit{transfers} the reputation of indicators. Each indicator is placed in a layer. If the number of reputation indicators is \textit{N}, there are \textit{N} layers. Each layer specifies its degree of importance to bootstrap its reputation. For example, the layer \textit{1} has the highest importance, where the layer \textit{N} has the lowest importance. %hence, if an indicator stays in the higher layer, the indicator is highly important in the reputation transfer process. 
A reputation indicator may stay in different layers based on the properties of a specific service. In Figure \ref{Fig3}, `provider' is placed in the first layer for service \textit{1}, it is placed in the second layer for service \textit{2}. In particular cases, some reputation-related indicators of one service may be unavailable to another service. For example, if a service does not belong to any community, the community information cannot be used to estimate its reputation. The number of layers, reputation indicators, the importance of the layers and the placement of indicators depends on the contexts of services.  We summarize the definitions of the major symbols used in the proposed approach in the Table \ref{tab:definition}.
\vspace{-3mm}
\begin{table}[htb]
  \caption{Terminologies}
  \vspace{-3mm}
  \label{tab:definition}
 %\resizebox{.5\textwidth}{!}{
\begin{tabular}{ |c|c| p{10cm} } 
 \hline
 \textbf{Symbol}&  \textbf{Meaning}\\ 
 \hline
 $Comp$ & \parbox{10cm} {A composition with $N$ component services}  \\ 
 \hline
 $s^{i}$ & \parbox{10cm} {The $i^{th}$ component service in $Comp$}  \\ 
 \hline
 $RI(s^{i})$  & \parbox{10cm}{the reputation-related indicators for the $s^{i}$ service} \\ 
  \hline
 $T(RI(s^{i}))$ & \parbox{10cm}{Types of reputation indicators for the $s^{i}$ service} \\ 
  \hline
 $NR(RI(s^{i}))$ & \parbox{10cm} {Normalised ratings of the indicator $RI(s^{i})$} \\
 \hline
 $importance_{f}$ & \parbox{10cm}{The importance of the indicator $f$} \\
  \hline
  $Lvl$ & \parbox{10cm}{The number of reputation classes or levels} \\
  \hline
  $\mathbb{L}$ & \parbox{10cm}{Feature matrix for model training} \\
  \hline
  $DFR^{i}$ & \parbox{10cm}{The Decision Tree Forest for the $i^{th}$ sample in the model training} \\
  \hline
  $\theta$ & \parbox{10cm}{A decision tree in the $DFR$}\\
  \hline
  $F$ & \parbox{10cm}{The forest matrix}\\
  \hline
  $SIG(x)$ & \parbox{10cm}{The sigmoid activation function of the neuron $x$}\\
  \hline
  $CRi$ & \parbox{10cm}{ The composite reputation for the $i^{th}$ sample in the model training} \\
  \hline
  $ac$ & \parbox{10cm}{Accuracy in model testing} \\
  \hline
  $bp$ & \parbox{10cm}{Bootstrapping probability, i.e., the confidence of the prediction} \\
  \hline
\end{tabular}
%}
\vspace{-4mm}
\end{table}
%particular cases. In the next section, we present a data-driven bootstrapping approach to determine the order of layers to predict the reputation of a new service. 

\subsection{Evaluating The Quality of Reputation Indicators}
%As explained in Section 3, the use layer is put in the first place of the framework for direct reputation evaluation.
%The reputation of indicators are transferred to bootstrap the reputation of the service in the proposed layered-based framework (Fig. \ref{Fig3}). Hence, we require to evaluate the reputation of the indicators at the first place. In the motivation scenario, if we want to bootstrap the reputation of the repository 'mopidy/mopidy' in GitHub, we need to first evaluate the reputation indicators, e.g., the quality of the contributors, project owner, and its community. 
%
%We assume that 
The quality of reputation indicators is evaluated by \textit{aggregating} consumers' feedback or ratings during a particular period. For example, the repository `mopidy/mopidy' in GitHub has 106 active contributors and each contributor is ranked or starred by fellow contributors. Similarly, the owner of the repository is `mopidy' which also gets a star value from the users of all of its repositories. Figure \ref{Fig3} depicts the importance of consumers' direct feedback in the layer-based reputation transfer. Note that, \textit{we do not consider an indicator in the bootstrapping process which does not have enough direct feedback to evaluate its quality}. %The lack of users' ratings create the \textit{recursive} problem of bootstrapping the quality of the reputation indicators which is a \textit{separate} problem itself.

Various methods are proposed to effectively aggregate the consumers' feedback or ratings which is the true reflection of the services' performance \cite{wang2011two,liu2011multi}. The rating is usually a quantitative value ranging from a \textit{minimum} to a \textit{maximum} value which is defined by the service provider. For example, ratings in IMDB\footnote{\url{https://www.imdb.com}} ranges from 1 (minimum) to 10 (maximum). Different factors such as the rater's credibility, past rating history, personal preferences, and temporal sensitivity are considered to create the true reflection of a rating in \cite{cho2009q}. As identifying the quality of the users' ratings is not the main focus of this paper, we consider only the \textit{rater's credibility} in aggregating the consumers' feedback. The rater's credibility, $c^{j}$ is a normalized value in the range [0,1]. Here, $c^{j} =0$ means `no credibility' and $c^{j} =1$ means the `highest credibility'. If \textit{N} consumers give ratings to a reputation indicator $RI^{i}$, the aggregated normalized rating $NR(RI^{i})$ is computed using the each consumer's ratings, $r^{j}$ and credibility, $c^{j} \rightarrow [0,1]$ in Equation \ref{nrating}. The value range of $NR(RI^{i})$ is [0,1].
\begin{equation}
\label{nrating}
\text{Normalized rating, } NR(RI^{i}) = \frac{\sum_{j}^{N} r^{j} \times c^{j}}{\sum_{j}^{N} r^{j}}.
\end{equation}  

The credibility of a rater should be assessed based on the \textit{context} of the application. In online shopping services such as eBay and Amazon, the credibility of a consumer is validated by the financial transactions of the consumers, i.e., whether the user is a true consumer of the service and does not have any malicious intentions. As we focus on the composite services in GitHub, financial transactions are not available. However, users often form a \textit{connected network} in open platforms such as GitHub, online forums and social networks \cite{kumar2010structure}. Figure \ref{Fig4} presents a peer to peer direct assessment among users.  If a user gathers more ratings from others, the credibility score of the user gets relatively higher. The intuition is that the peers recognize the expertise of the higher credible user. For example, consumer 1 in Figure \ref{Fig4} has higher credibility score (CS: 0.9) as it gets top ratings from all other consumers in its peer network. A rating given by a user who has a higher credibility score should have more weight than a rating given by a user who has a lower credibility score. Given a network of users, we compute the user credibility of giving ratings using the classic PageRank \cite{page1999pagerank}, which is a well-known approach to identify the importance of Web pages. The PageRank of a web page is defined recursively using the PageRank metric of all incoming links of pages. A page that is linked to by many pages with a higher PageRank receives a high rank itself. We choose the damping factor, d=0.85 which is generally used with the classic PageRank algorithm \cite{page1999pagerank}. A service platform may have its own rating systems. For example, eBay has a proprietary rating calculation algorithm for service providers. In such a case, we can directly use the provided \textit{credible} ratings to bootstrap the reputation of the new services.

% \begin{figure}[t!]
% 	\centering
% 	\begin{center}
% 		\scalebox{.85}{\includegraphics{figures/indicators.pdf}}
% 		%\includegraphics[width=.8\textwidth]{figures/indicators.pdf}
% 	\end{center}
% 	\vspace{-0.1in}
% 	\caption{Credibility-based peer to peer direct assessment}
% 	\vspace{0in}
% 	\label{Fig4}
% 	%\vspace{-0.2in}
% \end{figure}

\subsection{Data-driven Reputation Bootstrapping}
\label{atomicboot}
%The reputation bootstrapping process for a new service is to use its reputation-related indicators modeled in the layer-based framework to estimate its future direct reputation. 
We assume that there exists a \textit{history} of reputation-related indicators of different services and their observed direct reputation (feedback from consumers) which is normalized by an expert considering the qualitative nature of the indicators. \textit{The normalization process on each indicator is out of the focus of this paper.} For example, reputation indicators in a GitHub repository is already normalized in GitHub insight. As reputation is mostly qualitative in nature, the normalization process is intrinsically a human activity that requires domain knowledge to interpret the complex semantic meaning in the data transformation process. Furthermore, as the qualitative nature of services are highly context-aware, it is not realistic to apply a unique normalization process in every context. For example, the ratings of a travel service could not be compared to the ratings of a medical service as a slight downgrade in medical services may be critical and may impact highly on its ratings. Hence, we consider human in the loop to ensure the consistency, i.e., comparing the indicators in the right way. The ``age of a repository'' is one of the reputation indicators in the training set that indicates the temporal influence of other reputation related indicators such as watchers, commits, and stars. Our target is to leverage the history and to apply supervised machine learning techniques \cite{liaw2002classification,kong2018deep} to find the relative importance of reputation indicators. We collect the history with \textit{N} instance or samples of an observed atomic service and build a feature matrix $[T(RI^{i}), NR(RI^{i})]$ where $i \in [1, K]$ for the model training. Each service instance or sample represents the information on reputation-related indicators and the \textit{observed} direct reputation. %Fig. \ref{Figinstance} depicts \textit{N} instances in a history. 
$T(RI^{i})$ represents the type of the indicator $RI^{i}$ (categorical variable) and $NR(RI^{i}) \rightarrow [0,1]$ denotes the normalized rating of $RI^{i}$ for an instance. If $\mathbb{L}$ contains all possible reputation-related indicators, an instance can have a maximum  $|\mathbb{L}|$ numbers of reputation indicators. However, all reputation indicators from one service may not be applicable to new services. Here, the instance may contain a subset of reputation indicators, $L \subseteq \mathbb{L}$. %In Fig. \ref{Figinstance}, instance 1 has no entry for reputation indicator $RI^{i}$.  
%\begin{figure}[htb]
%	\centering
%	\begin{center}
%		\includegraphics[width=.48\textwidth,height=.12\textwidth]{figures/instances.pdf}
%	\end{center}
%	\vspace{-0.30in}
%	\caption{Features and instances in a historical data for reputation bootstrapping}
%	\vspace{0in}
%	\label{Figinstance}
%	%\vspace{-0.2in}
%\end{figure}
%
%We also consider the types of the reputation indicators in the history.  
We may not get all the types of reputation indicator for a new service. For example, the type of contributor in a repository of a GitHub may be any combination of `freelancer', `employed', `student' and so on. In certain cases, such information may not be explicitly mentioned or retrieved from GitHub. We denote such types as unknown or \textit{`U'}. %In Fig. \ref{Figinstance}, the type of $RI^{2}$ is unknown or 'U'.

The observed reputation of a service has two types: a) \textit{quantitative}, and b) \textit{qualitative}. The quantitative reputation of a service $S^{i}$ is the normalized rating $NR(S^{i})\rightarrow [0,1]$ using Equation \ref{nrating}. The qualitative reputation has semantic interpretation such as `high', `moderate', `low' and so on. If there exists $Lvl$ reputation levels, the range of $i^{th}$ level, $Lvl^{i} \rightarrow [0,1]$ is computed in Equation \ref{qualidata}. %Fig. \ref{Figinstance} represents qualitative observed reputation of a service in the history.
\begin{equation}
\label{qualidata}
Lvl^{i} = [(i-1) \times \frac{1}{Lvl}, i \times \frac{1}{Lvl}], \; i \;\epsilon\; [1, Lvl].
\end{equation}     
 
\textit{We mainly focus on bootstrapping the qualitative reputation from historical records.} We consider the bootstrapping process as supervised learning for classification. Our objective is to learn a function $R(L) = \hat{r}$ to approximate $\hat{r} \approx r$, where $\hat{r}$ is the bootstrapped reputation of a new service, $r$ is the direct reputation and $L$ is the set of reputation-related indicators. We also determine the importance of indicators in reputation bootstrapping, i.e., $I(\mathbb{L}) = \overset{\rightharpoonup}{i}$ outputs a vector $\overset{\rightharpoonup}{i}$ which contains the importance value of every indicator in $\mathbb{L}$. A supervised learning method learns the functions $R$ and $I$ using the collected features in every layer of $\mathbb{L}$ in history.

\subsubsection{Learning $R$ and $l$ Functions}
There exist various machine learning approaches for supervised classification or prediction such as Decision Tree, Random Decision Forest, Support Vector Machine, Logistic Regression and Neural Networks (Multilayer Perceptron) \cite{liaw2002classification,kong2018deep,silver2016mastering}. We choose Random Decision Forest \cite{liaw2002classification} as the preferred learning approach in bootstrapping reputation for a single service. Random Decision Forest is an ensemble learning algorithm based on the decision tree. A Support-Vector Machine (SVM) constructs a hyperplane or set of hyperplanes in a high or infinite-dimensional space, which can be used for classification, regression, or other tasks like outlier detection \cite{spinello2008human}. Deep Neural Network (DNN) is a class of machine learning algorithms that uses multiple layers to progressively extract higher level features from the training dataset in the supervised learning \cite{silver2016mastering}. The key reasons to choose Random Forest over other approaches are:
\begin{itemize}
[topsep=0pt,parsep=0pt,partopsep=0pt,leftmargin=10pt,labelwidth=6pt,labelsep=4pt]
\item The feature matrix in $\mathbb{L}$ is sparse rather than dense. The reputation indicators which are applicable for a certain type of service, may not be applicable to another set of services. Hence, $L \subseteq \mathbb{L}$ creates the sparse feature matrix in the training dataset. Random Forest naturally handles various cases of incomplete features through a feature bagging process \cite{liaw2002classification}. In contrast, multi-class SVM usually requires data prepossessing, e.g., clustering to identify incomplete features before the multi-class classification \cite{spinello2008human}. Note that DNN is also a natural fit to handle incomplete sparse feature matrix \cite{silver2016mastering}.     
\item The history of all single services consists of a large number of small clusters of services where the same set of indicators are applied. These subsets are not large enough to be effectively trained by deep neural networks and other machine learning approaches \cite{silver2016mastering}. Note that DNN is usually preferred in classification problems when the dataset is large and feature matrix is unstructured. In contrast, small clusters in a large dataset are naturally fit to small random decision forests due to the semi-structured nature of the dataset \cite{kontschieder2015deep}. The training of all the possible $L \subseteq \mathbb{L}$ can be done in a single step.

\item One of the objectives is to learn $l$ function, i.e., learning the importance of the indicators in various cases. The Random Forest inherently learns the feature importance \cite{liaw2002classification}. The importance value of every feature in each layer of $\mathbb{L}$ can be aggregated to compute the importance of each reputation-related indicator. 
\end{itemize}  

%We modify the standard random forest algorithm to address the various cases of  $L \subseteq \mathbb{L}$. We consider a feature based bagging process along with the standard bootstrapped bagging as follows:
We apply the standard random forest algorithm to bootstrap the reputation. We modify the bagging process of the standard random forest algorithm to address the various cases of  $L \subseteq \mathbb{L}$. We consider a feature-based bagging process, i.e., Reputation Indicator Bagging (vertical sampling) along with the standard bootstrapped bagging as follows. 
\begin{itemize}
[topsep=0pt,parsep=0pt,partopsep=0pt,leftmargin=10pt,labelwidth=6pt,labelsep=4pt]
\item Bootstrap Bagging (horizontal sampling): Let us assume that the original learning set of $\mathbb{L}$ are composed of $P$ samples or instances. We first generate $K$ learning sets, $L\_Full^{k}$, which is composed of $q$ samples where $q \leq P$. The samples are randomly selected (uniform sampling) with replacement from $\mathbb{L}$. The intuitive reason for horizontal sampling is that it restricts over-fitting on the training data by reducing the variance of the prediction function \cite{kotsiantis2011combining}.

\item Reputation Indicator Bagging (vertical sampling): Each of the $L\_Full^{k}$ are partitioned in several smaller learning sets,  $L\_Partial^{k}$ based on the subset of reputation indicators $L^{i} \subseteq \mathbb{L}$. The feature matrix $[T(RI^{i}), NR(RI^{i})]$ could be partitioned into two subsets: a) \textit{even}: $\{T(R^{1}), NR(R^{1}),T(R^{3}), NR\\(R^{3}),... ,T(R^{i}), NR(R^{i}), R \}$ and \textit{odd}: $\{T(R^{2}), NR(R^{2}), T(R^{4}), NR(R^{4}),...,T(R^{i+1}), NR(R^{i+1}), R \}$. The intuitive reason for such vertical sampling is that ``\textit{if a subset of $L$ are only related to the reputation of a service, the random forests based on the subset of $L$ should be sufficient for the bootstrapping.}'' 
%service can only be modeled in a particular $L$ according to its indicators, its future reputation should be predicted through the sub-forest that is built on $L$}". 
For example, if a new service has the information of its provider and similar services, It should be sufficient to bootstrap its reputation only based on a particular $L$ consisting of a provider layer and a similar service layer.
\end{itemize}  

Algorithm \ref{ag1} describes the random forest building approach for bootstrapping the reputation. Both the ``\textit{Bootstrap Bagging}'' and ``\textit{Reputation Indicator Bagging}'' are applied to create $k$ subsets of training data from the history. Each of the training subsets is applied to create decision trees using the CART (Classification And Regression Trees) algorithm \cite{kotsiantis2011combining}. The feature matrix $[N \times M]$ are traversed in $O(MN)$ where $N$ is number of sample rows and $M$ is number of features including the output variable (nested loop in lines 1 to 8 in Algorithm \ref{ag1}). For each iteration, the trees are fully grown and not pruned (as may be done in constructing a normal tree classifier) and form a forest (lines 5 to 7 in Algorithm \ref{ag1}). Existing related work strongly suggest that full grown trees over pruning in Random forest algorithm to be a better alternative \cite{breiman2001random,liaw2002classification}. Although pruning is a suitable approach used in decision trees to reduce over-fitting, some of the potential over-fitting is mitigated in Random Forest. As the proposed random forests training use bootstrap bagging along with random process of reputation indicator bagging, the correlation between the trees (or weak learners) would be low. Note that, the complexity of building a decision tree is $O(MNlog(N))$ in python scikit-learn\footnote{\url{https://scikit-learn.org/stable/modules/tree.html}}. All decision trees are used to predict the reputation of the new service. The average vote of all trees is reported as the Random Forest prediction. Hence, the complexity of Algorithm \ref{ag1} is $O(M^{2}N^{2}log(N))$. Note, different aggregation functions, e.g., average, median, mode, majority voting, and weighted aggregation could be used to make prediction in Random Forest \cite{liaw2002classification}. The average voting of all trees is the default option in the standard Random Forest and intuitively should reduce the overall variance \cite{breiman2001random}. The random sampling with full-grown trees generally induces both under-bias as often as over-bias trees. As a result, it is expected that the average will cancel the over and under bias of each tree and thus reduce the reduce the variance in each tree \cite{ghosal2020boosting}.

%\vspace{-.1in}
\renewcommand{\algorithmicrequire}{ \textbf{Input:}}
\renewcommand{\algorithmicensure}{ \textbf{Output:}}
\begin{algorithm}[t!]
	\caption{Random Forest Building for Reputation Bootstrapping}
	\label{ag1}
	\scriptsize
	\begin{algorithmic}[1]
		\REQUIRE ~~\\
		$\mathbb{L}$: the set of all reputation indicators. \\
		The training set $N$ containing $P$ samples;\\
		$\{L_i\}$: all possible subsets of $\mathbb{L}$, where $L_i \subset \mathbb{L}$\\
		\ENSURE
		The subsets of decision trees that form a forest.
		
		\FOR{$i = 1 \dots N$ }
		\STATE Create $L\_Full^{i}$ training set using Bootstrap Bagging (Horizontal sampling) with replacements.
		\STATE Build an unpruned decision tree $tr^t_i$ based on $L\_Full^{i}$.
		\FOR{$j = 1 \dots M$}
		\STATE Choose a random $L_i \in \{L_i\}$.
		\STATE Create $L\_partial^{i}_{j}$ training set using  Reputation Indicator Bagging (vertical sampling) with replacements.
		\STATE Build an unpruned decision tree $ftr^t_{ij}$ based on $L\_partial^{i}_{j}$
		\ENDFOR
		\ENDFOR
		\RETURN a decision forest ($DFR$) based on the combination of  $tr^t_i$ and $ftr^t_{ij}$.
	\end{algorithmic}
\end{algorithm}
%\vspace{-.1in}

We apply classification trees in our proposed approach rather than regression trees. The classes represent different levels of reputations, e.g., ``Very high'', ``High'', ``Moderate'', ``Low'', and ``Very low''. The reason is that a consumer in practice usually cares qualitative reputation classes, rather than a quantitative reputation value, which is hard to be interpreted in the term of \textit{reputation} naturally.

 \subsubsection{Finding The Importance Of An Indicator}
 \label{relativeimportance}
As random forest approach is based on decision trees, it inherently calculates the importance of reputation-related indicators for an atomic service \cite{qu2017confidence}. It is important to identify influential indicators. Some key reasons are as follows:
\begin{itemize}
[topsep=0pt,parsep=0pt,partopsep=0pt,leftmargin=10pt,labelwidth=6pt,labelsep=4pt]
\item \textit{Dimension reduction}: The less important indicators may be considered \textit{redundant} and noisy. They could be omitted to improve the accuracy of the prediction. 
\item \textit{Computational efficiency}: Once the dimension gets reduced, the learning rate and computation efficiency of the prediction should improve significantly.     
\end{itemize}

There exist different approaches to determine the importance of variables or features in random forest \cite{liaw2002classification}. We apply Mean Decrease Accuracy (MDA) and Mean Decrease Information Gain (MDIG) \cite{kotsiantis2011combining} approaches to find the importance of reputation-related indicators.
\begin{itemize}
[topsep=0pt,parsep=0pt,partopsep=0pt,leftmargin=10pt,labelwidth=6pt,labelsep=4pt]
\item \textit{Mean Decrease Accuracy (MDA)}: As we create a forest ($DFR$) of decision trees using bootstrap bagging process with replacement in Algorithm \ref{ag1}, there exist out-of-bag (\textit{oob}) samples or instances where no decision tree is created. The oob samples are tested with each constructed decision tree and the prediction accuracy, i.e., the rate of correct classifications is calculated using Precision and Recall. Our focus is to apply model-based feature importance methods \cite{parr2020nonparametric} to rank and eliminate weak predictors during model development. The general approach is to tweak the model's parameters/indicators and measure the tweak’s effect on model prediction accuracy. The simplest approach is \textit{drop-column importance} \cite{gregorutti2017correlation}, which defines the importance of an indicator by calculating the difference in prediction accuracy between a model with all features (the baseline) and a model with the indicator removed. As the indicator is removed, the model requires retraining. To avoid retraining, the Random Forest algorithm use the Permutation-importance method \cite{breiman2001random}. Here, the values for a certain indicator are randomly permuted for all oob samples. The prediction accuracy is computed using the permuted samples for all trees and compared against the original accuracy. If the accuracy remains similar, the null hypothesis is that the importance of the indicator is lower as the changes in its values are not affecting the quality of the learning. A significant difference states that the indicator has higher importance. Finally, the importance is calculated using the standard deviation of the prediction accuracy. For example, let us assume,  $DFR$ is constructed from 100 training samples which include 5 reputation-indicators (i.e., provider, community, similar services, age of service, and customer service) to classify 3 reputation classes (i.e., high, medium, and low). There exist 20 oob samples to test the accuracy of the $DFR$. In the first test, 15 out of 20 samples can accurately predict the reputation classes, hence the accuracy is .80. Next, the values of only the `provider' are randomly shuffled in the oob test matrix and the accuracy decreases 50\% to .40. However, if only the values of `community' are randomly shuffled, the accuracy decreases 10\% to .72. Let us assume, similar results are found in all the trees in the forest. It shows that the feature `provider' has higher MDA and is more important than `community' as it has higher effect in the learning accuracy. We will explore the conditional permutation importance method \cite{parr2020nonparametric} in the future work.          

\item \textit{Mean Decrease Confusion Degree (MDCD)}: Each decision tree in $DFR$ has two types of nodes: a) decision nodes, and b) leaf nodes. The leaf nodes contain the reputation values of the service and the decision nodes represent the condition on related indicators. Each decisions node is split into two child nodes based on a single indicator in order to make similar samples stay in the same node, i.e., reducing the `impurity' \cite{kotsiantis2011combining}. The impurity values are typically computed through \textit{Gini impurity} or \textit{information gain (IG)} \cite{liaw2002classification}. The training process of a decision tree is to determine how quickly each indicator can reduce sample impurity in a decision tree. If $IG$ denotes impurity values on information gain, $NS$ denotes the number of samples in a node, the importance of an indicator (value ranges in [0,1]) at the $i^{th}$ node with $r$ right child node and $l$ left child node can be computed using Equation \ref{Eq2}: 
\begin{equation}
	\small
	\label{Eq2}
	\text{importance} = IG_n \times NS_n - IG_l \times NS_l - IG_r \times NS_r.
\end{equation}

We compute the global importance of indicators in a decision forest as the average of the importance of indicators computed in every single tree. Let $importance_f$ denote the importance of a indicator $f$ in $FR$ built on $\mathbb{L}$;
$importance_f^i \rightarrow [0,1]$ denotes the importance of a feature~$i$ belonging to $f$.
The importance of a indicator $f$, $importance_f$ is the sum of the importance of all the features, $n$, $importance_f^i$ belonging to $f$ as follows:
\begin{equation}
	\small
	\label{Eq3}
	\text{importance}_f = \frac{\sum_i^{n} \text{importance}_f^i}{n}.
\end{equation}

Based on the importance score, we order the layers of $\mathbb{L}$. The bootstrapped reputation based on higher importance indicators can be considered more reliable. For example, let us assume the $IG$ of an intermediate node in a $DFR$ is .54 which is the probability of incorrectly classifying a randomly chosen element in the training dataset if it were randomly labelled according to the class distribution in the dataset. The left branch of the node is split using the feature `provider' and the right branch is split using the feature `community'. The IG of the left child is .20 and IG of the right child is .48. It implies that `provider' has higher decrease in IG to accurately split the data points. Let us assume the MDCD of `provider' and `community' is 100.3 and 20.8 respectively which are calculated over all $DFR$ in the forest. It implies that `provider' is a significant feature in comparison with `community' in the learning model.              
\end{itemize}

\section{Reputation Bootstrapping For Composite Services}
Let us assume that a composite service, $Comp$ has $N$ component services as $Comp = \{s^{1}, s^{2},\dots, s^{N}\}$. Each service $s^{i}$ has a set of reputation indicators, $RI(s^{i})$ and corresponding values, i.e., type of the indicators $T(RI(s^{i}))$ and normalized ratings of the indicators $NR(RI(s^{i}))$. Our target is to predict the future direct feedback or ratings of the composite service, i.e., the reputation of the composition.

\subsection{Topology Free Reputation Bootstrapping (TFRB)}
TFRB is a naive approach where we consider only the component services but discard the composition topology. The intuition behind composition topology free reputation bootstrapping is that ``the composite reputation is not dependent on the directions of service invocation, rather dependent on the aggregated reputation indicators of the component service''. \textit{We assume that there is a historical data on composite services, i.e., information on the various reputation indicators and the observed composite reputation value.} Our objective is to learn relative importance between the reputation influence across component services.

If we do not consider the topology, the reputation bootstrapping for composite services could be treated as reputation bootstrapping through the \textit{aggregation} of several atomic services. The individual history of each component services could be appended together in a single history. Figure \ref{Figtopologyfree} depicts a history of compositions ($N$ component services). Note that, if our objective is to find the composite reputation of $m$ component services, we should build a feature matrix containing the reputation of existing composite services where the number of component services is $m$. In Figure \ref{Figtopologyfree}, for each composition samples or instances, we have the type function $T()$ and normalized rating $NR()$ described in Section \ref{atomicboot} of each component services in the history. The observed qualitative composite reputation, i.e., direct feedback from the consumers are collected for the supervised training using Equation \ref{qualidata}. Similar to Section \ref{atomicboot}, we consider the bootstrapping of composite reputation as a \textit{classification learning} rather than regression. 

% \begin{figure}[htb]
% \centering
% 		\includegraphics[width=.85\textwidth]{figures/Instances2.pdf}
% 	\vspace{-0.2in}
% 	\caption{Aggregating component service features in a single feature matrix}
% % 	%\vspace{0in}
% 	\label{Figtopologyfree}
% % 	\vspace{-0.1in}
%  \end{figure}

Once the feature matrix is generated from the history, we apply the Algorithm \ref{ag1} to generate the random decision trees based on the composition samples. The relative importance of the reputation indicators are calculated using the proposed approach in Section \ref{relativeimportance}. Finally, the reputation of a new composition is estimated by the following two steps:
\begin{enumerate}
[topsep=0pt,parsep=0pt,partopsep=0pt,leftmargin=10pt,labelwidth=6pt,labelsep=4pt]
\item For each generated decision forest tree $(DFR^{i})$, traverse the tree based on the the type function $T()$ and normalized rating $NR()$ of each reputation-related indicators and estimate the qualitative composite reputation, $CR{i}$.
\item The final prediction is chosen based on majority voting \cite{liaw2002classification}. If $CR{i}$ is estimated as the leaf node in the tree traversal for $(\frac{N}{2}+1)$ decision forest trees, we output the $CR{i}$ as the composite reputation. The key reason to consider majority voting is that it eliminates the bias of each full length-no pruning decision, i.e., the over and under biases. It reduces the variance in each tree and so the overall variance should be reduced as well.     
\end{enumerate}     
\vspace{-.15in}
\subsection{Topology-aware Reputation Bootstrapping} 
\begin{figure}[t!]
%\vspace{-.2in}
\begin{minipage}{.5\textwidth}
\centering
\includegraphics[width=1.1\textwidth,height=.4\textwidth]{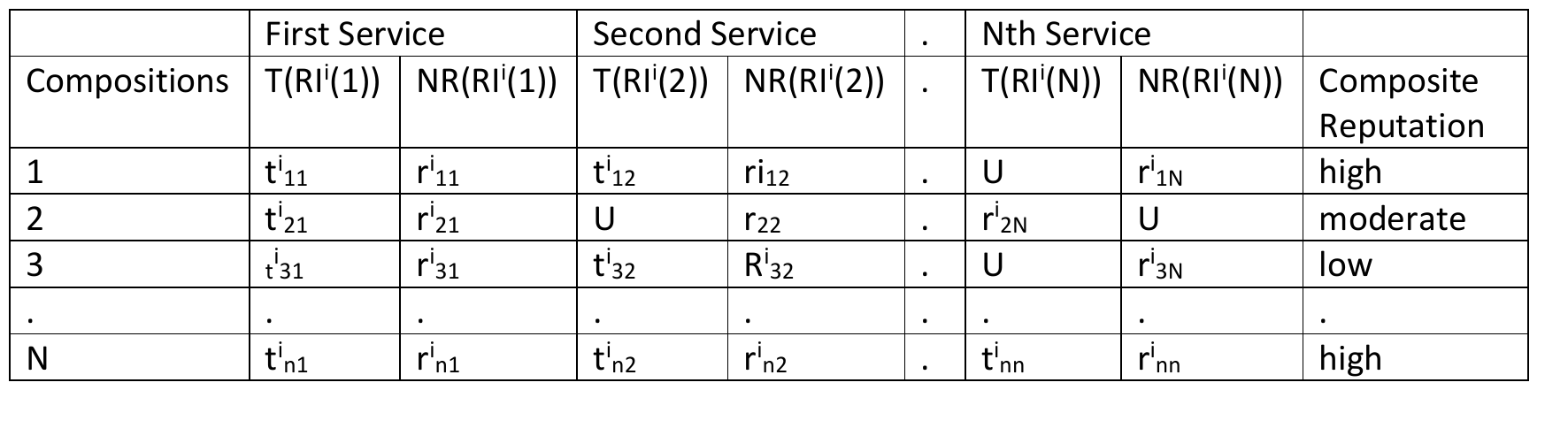}
	\vspace{-0.2in}
\caption{Aggregating component service features in a single feature matrix}
	\label{Figtopologyfree}
\end{minipage}
\hspace{.4cm}
%\vspace{-0.2in}
\begin{minipage}{.45\textwidth}
	\centering
	\begin{center}
	    \vspace{-0.4in}
		\includegraphics[width=.88\textwidth,height=.32\textwidth]{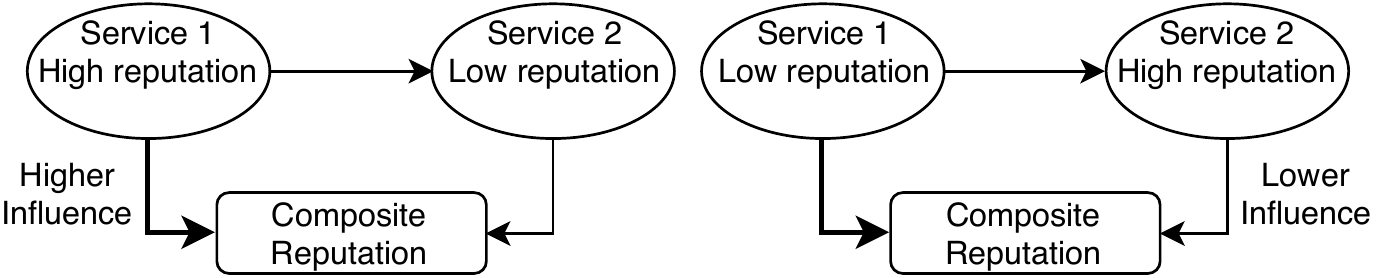}
	\end{center}
	%\vspace{-0.30in}
	\caption{Reputation influence of the service invocation direction}
	\vspace{-.5in}
	\label{Figrinflu}
	%\vspace{-0.2in}

\end{minipage} 
\vspace{-.2in}
\end{figure}

In the naive approach, i.e., topology free bootstrapping, we do not consider the influence of a composition topology on the composite reputation. However, the composition topology and the directions among service invocations may have a greater influence on the reputation of the composition. We term such phenomena as `\textit{reputation influence}'. The reputation influence represents that if a highly reputed service invokes a new service, the composite reputation may also be treated at a higher level. The rationale is that highly reputed services are usually cautious about their reputation and implement rigorous QoS testing in the process of invoking new services. For example, `Poli' is a new payment service. When Amazon starts supporting `Poli' in its payment platform, the reputation of Amazon influences the reputation of `Poli'. When a lower reputed service invokes a highly reputed service in a composition, it may not have a higher influence on estimating the reputation of the composition. For example, a new VPN service may invoke the highly reputed `Paypal' service for payment. It does not translate that the reputation of the VPN service should be higher as there may be performance and security issue. Hence, the reputation influence is \textit{directional}. Figure \ref{Figrinflu} describes the influence of service invocation direction in the composite reputation. The reputation influence also depends on the different layers of the reputation-related factors.  We need to consider the service topology in the reputation bootstrapping.

% \begin{figure}[htb]
% 	\centering
% 	\begin{center}
% 		\includegraphics[width=.6\textwidth,height=.15\textwidth]{figures/reputationinfluence.pdf}
% 	\end{center}
% 	%\vspace{-0.30in}
% 	\caption{Reputation influence of the service invocation direction}
% 	\vspace{0in}
% 	\label{Figrinflu}
% 	%\vspace{-0.2in}
% \end{figure}

There exists different topology patterns: a) \textit{sequential}, b) \textit{parallel or conditional} and c) \textit{hybrid} \cite{4279708}. The hybrid pattern consists of both sequential and parallel component. Lets us assume, the new composite service has a composition pattern $Pat^{i}$ with $m$ component services. We collect all the composition of pattern $Pat^{j}$ with $m$ component services from the previous compositions. This will form the training dataset to build a machine learning model for $Pat^{i}$.
      
\subsubsection{Forest Deep Neural Network (fDNN) for Composite Reputation Bootstrapping}
\label{sec:fDNNcom}
We model the reputation bootstrapping of composite services using Forest Deep Neural Network (fDNN) \cite{kong2018deep}.  The fDNN is a state-of-art machine learning technique consisting of two parts. The forest part serves as a feature detector to learn sparse representations from raw inputs with the supervision of training outcomes, and the DNN part serves as a learner to predict outcomes with the new feature representations. The fDNN is naturally fit to classify composite reputations with higher accuracy as key features from both Random Forest and DNN are incorporated in the fDNN. The key reasons to choose fDNN over Random forest and DNN are:

\begin{itemize}
\item Random forest is an excellent tool to learn feature representations given its robust classification power and easily interpretable learning mechanism with semi-structured feature matrix \cite{liaw2002classification}. While the reputation indicators are semi-structured for atomic services, the feature matrix of composite services is usually sparse and unstructured due to the complex unknown correlation among reputation indicators from different services, various composition topology and unknown reputation influence. Hence, Random Forest as a single training model may not be suitable for the highly accurate classification of composite reputation. In contrast, the training of fDNN classifier consists of two stages. In the first stage, training data including labels are used to fit the forest. Predictions from each tree in the forest for all instances are then fed into the fully-connected DNN, for training in the second stage which considers the unstructured feature matrix.

\item The history of all composite services consists of a large number of small clusters of composition where the same set of indicators and same topology are applied. Given a new composition (topology pattern $Pat^{i}$ and $m$ component services), we may not gather a large number of similar compositions in history. We will end up a large number of reputation indicators (columns in the feature matrix) but a smaller number of samples (rows in the feature matrix). These subsets are not large enough to be effectively trained by deep neural networks and other machine learning approaches \cite{hunter2012selection}. As the first stage of fDNN creates large feature representations of forests, the DNN in the second stage should be able to model relationships among reputation-indicators of atomic services in various composition topology.    

\end{itemize}

%We can learn the reputation influence among different indicators in various layers using deep learning techniques such as Deep Neural Network (DNN), Recurrent Neural Network (RNN), and Forest Deep Neural Network (fDNN) \cite{silver2016mastering}. 

%The DNN is usually highly efficient when the training dataset is relatively large \cite{hunter2012selection}. However, given a new composition (topology pattern $Pat^{i}$ and $m$ component services), we may not gather a large number of similar composition in history. As a result, we will end up a large number of reputation indicators (columns in the feature matrix) but a smaller number of samples (rows in the feature matrix). It may affect both the training and testing accuracy of the DNN model.  Moreover, the existence of complex unknown correlation among reputation indicators from different services adds extra complexity for accurate reputation bootstrapping.

%We model the reputation bootstrapping of composite services using Forest Deep Neural Network (fDNN) \cite{kong2018deep}. The fDNN is a state-of-art machine learning technique that builds feature importance on top of DNN classifiers to achieve correlations among features with fewer samples in comparison to DNNs \cite{kong2018deep}. 
We incorporate reputation indicators of component services in a fDNN as follows:
\begin{itemize}
[topsep=0pt,parsep=0pt,partopsep=0pt,leftmargin=10pt,labelwidth=6pt,labelsep=4pt]
\item \textit{Reputation indicator importance detection}: The forest part of the fDNN serves as a detector to learn sparse representations of reputation indicator importance from supervised training samples. Random forest is the chosen approach to generate the sparse representation as it is an ensemble of independent decision trees \cite{kontschieder2015deep}. Algorithm \ref{ag1} returns a set of $M$ decision forest trees for $N$ samples and the corresponding observation, as $DFR = \{\theta_{1}, \theta_{2},....,\theta_{M}\}$.

\item \textit{Reputation prediction DNN model}: The DNN part in a fDNN learns to predict outcomes with the new sparse representations of random forests. Given a $DFR$, we generate the forest matrix, $F = \{f_{1}, f_{2},....,f_{m}\}$ where each $f_{i}$ is the prediction of a sample from a tree $\theta_{i}$ with an observation $x_{i}$. Each predicted $f_{i}$  serves as the new input features to be fed into the DNN.   
\end{itemize}

The DNN part is a standard deep neural network with $l$ hidden layer, an input layer, and an output layer.  Each hidden layer consists of a number of neurons. The input layer is formed with nodes by the forest matrix $F$. The nodes in the output layer are mapped with each class of the reputation (`high', `low', and etc.) in the outcome vector $r(Lvl)$ based on the qualitative $Lvl$ level using Equation \ref{qualidata}. Nodes and neurons in each layer are interconnected with links called \emph{synapses}. Each link is associated with a \emph{weight}, $W$ and a bias value $B$. The set of hidden neurons are denoted by $Z$. Usually, the number of hidden neurons in the ($|z| = h$) decreases from the input layer $(h_{i} > h_{i+1} )$. If $\gamma$ represents all the parameters in a DNN, we formally represent the DNN as predicting the outcome vector $r(Lvl)$ as the outcome probability, $Pr(r(Lvl)| F,\gamma)$ as follows:  
\begin{align}
&Z_{1} = \beta(F W_{in} + B_{in})
\hdots,
Z_{l} = \beta(Z_{l-1} W_{l-1} + B_{l-1})
\hdots,
Z_{out} = \beta(Z_{l}W_{l} + B_{l}) \\ \notag
&Pr(r(Lvl)| F,\gamma) = \alpha(Z_{out}W_{out} + B_{out}). \notag
\end{align}        

Here, $\beta$ is the sigmoid activation function for the neurons and $\alpha$  is the softmax function. The softmax function is used to transform the values of the output layer into the probability prediction \cite{silver2016mastering}. The Equation \ref{eq:nout} describes the calculation of the output of neuron $x$ on a link $(i,j)$ as $\mu_{ij}$.

\begin{align}
\label{eq:nout}
SIG(x)=\frac{1}{1+e^{-x}}, \;\;
\mu_{ij} = [Z_{i}^{(out)}]^{T} W_{j}^{(out)} +B_{i}.
\end{align}
        
\subsubsection{fDNN Chaining for Composition Topology}
A single fDNN does not consider the reputation influence which is generated by the service invocation directions in composition. For example, if a service $X$ invokes a service $Y$ and $Y$ requires service $Z$, ($X\rightarrow Y \rightarrow Z)$, the reputation indicators $Z$ have direct interactions with the indicators of $Y$, but have indirect interactions with the the indicators of $Y$, i.e.,  ($X\rightarrow [Y \rightarrow Z])$. Note that the indicators of $Y$ have direct interactions with both $X$ and $Y$, i.e.,  ($[X\rightarrow Y]\rightarrow [Y\rightarrow Z])$. A single fDNN only represents the direct invocations between service indicators. Hence, we need to chain several fDNN to capture both direct and indirect invocations. The chained fDNN facilitates reputation influence transfer for the composition.

\begin{figure}[t!]
\vspace{-0.2in}
	\centering
	\begin{center}
		\includegraphics[width=1\textwidth,height=.38\textwidth]{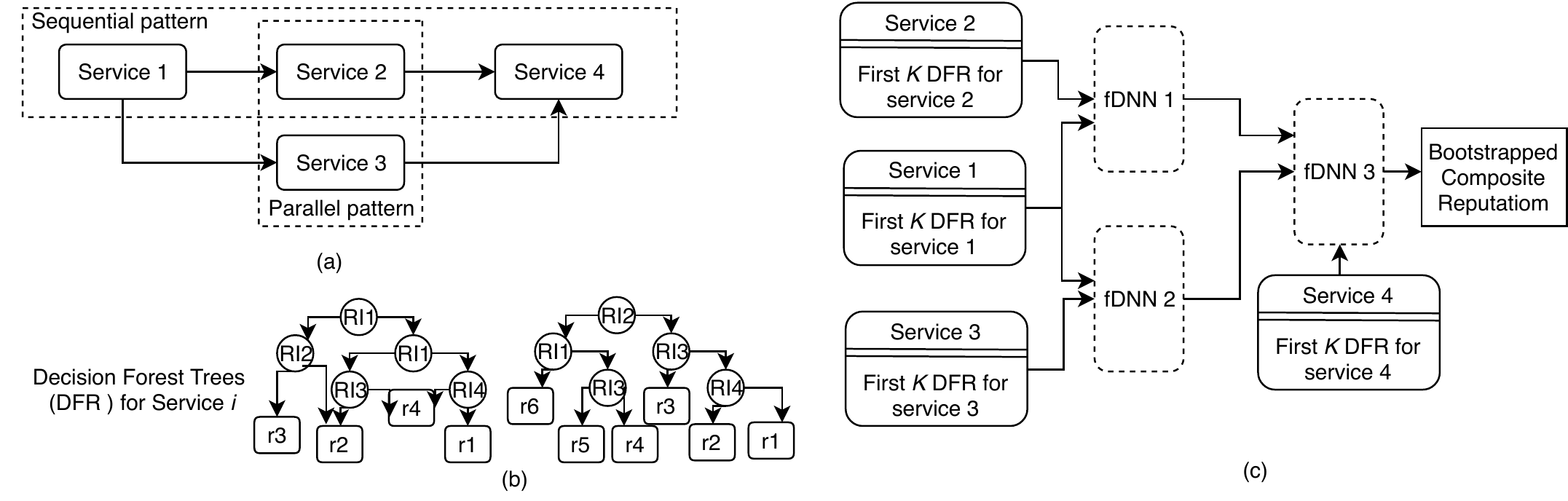}
	\end{center}
	\vspace{-0.2in}
	\caption{A fDNN chaining process example. (a) A simple hybrid topology, (b) Decision Forest Trees for a component service, and (c) fDNN chaining for the composition topology}
	%\vspace{0in}
	\label{Figfdnn}
	\vspace{-0.2in}
\end{figure}

Algorithm \ref{ag2} describes the chaining process of a generic topology. We assume that we have training data for the respective composite topology. The fDNNs are chained in a bottom-up manner. All the component services are ordered (highest to lowest) based on their invocation dependency depth. The dependency depth is calculated by the depth first search (DFS) tree traversing starting from the composed service. For example, Figure \ref{Figfdnn}(a), the dependency depths of `service 1', `service 2', `service 3', and `service 4' are 3, 2, 2, 1 respectively. For $M$ services in a composition, the complexity of the traversal is $O(M^{2})$ (the nested loop in lines 3 to 22 in Algorithm \ref{ag2}). Starting from the service with the highest depth, a new fDNN is created when there is a direct invocation between two component services (lines 7 to 9 in Algorithm \ref{ag2}). According to \cite{kong2018deep}, with $n$ training samples, $m$ features, $k$ hidden layers, each containing $h$ neurons, and $o$ output neurons, the time complexity of back propagation in fDNN is  $O(n \times m \times h^{k} \times o \times i)$ when $i$ is the number of iterations. If the input layers of the corresponding services are not created yet, we create random forests for each service and select top $k$ $DFR$ on the input layer (lines 24 to 28 in Algorithm \ref{ag2} and the complexity is $O(M^{2}N^{2}log(N))$ according to Section 2.3.1). If the input layer is already created, the input layer of the new fDNN is fed with the corresponding already created fDNN (lines 15 to 17 in Algorithm \ref{ag2}). The total complexity of Algorithm \ref{ag2} is $O(M^{2} \times n \times m \times h^{k} \times o \times i)+ O(M^{2}N^{2}log(N))$. we have training data for the respective composite topology. We will explore fDNN chaining with incomplete information in the future work.

Figure \ref{Figfdnn} illustrates the chaining process of fDNNs for a hybrid topology following Algorithm \ref{ag2}. Figure \ref{Figfdnn}(a) depicts a simple hybrid pattern with 4 services. Both sequential and parallel pattern exists in the hybrid pattern. Figure \ref{Figfdnn}(b) represents a sample random decision forests using Algorithm \ref{ag1} for each component service.  Each forest matrix (according to \ref{sec:fDNNcom}) will be used in the input layer of the fDNN. Figure \ref{Figfdnn}(c) shows the chaining of 3 fDNNs for the composition topology. The `service 1' has direct invocations with `service 2' and `service 3', but there is no direct invocation between `service 2' and `service 3'. Hence, one fDNN is modeled using forest matrices from both `service 1' and `service 2' as the input layers and another fDNN is modeled using forest matrices from both `service 1' and `service 3' as the input layers. Note that, both `service 2' and `service 3' have direct invocations with `service 4'. As forest matrices of both `service 2' and `service 3' are modeled as input layers, the third fDNN is constructed based on the two previous fDNNs and the forest matrix of `service 4'. The output layer is mapped with the reputation classes.

\subsubsection{The Chained fDNN Model Training}
The goal of the training phase is to find a set of weights that, if substituted in the mapping functions along with the inputs, maximizes the accuracy of evaluating the observed reputation. The DNN uses an optimizer to adjust the weights to the provided sample data accordingly. We use Adam Optimizer \cite{du2018towards}, however, any other optimizer can be used with the algorithm. The task of the optimizer is to minimize the cross-entropy loss function \cite{du2018towards}. If $\gamma$ represents the parameters in the fDNN, the actual composite reputation is $r$ and the fitted reputation is $\bar{r}$, cross-entropy loss function \cite{kong2018deep}, $Q$ is denoted as follows:
\begin{equation}
Q(\gamma) = -\frac{1}{n}\sum_{i =1}^{n}{r(Lvl) \;log\;\bar{r}+ (1- r(Lvl)) log(1 - \bar{r})}.
\end{equation}   

%The training accuracy of the fDNN depends on the parameters for the reputation indicator importance detection part and the DNN part for prediction. 
The random forest parameters in a fDNN include the information gain value, depth of the tree, the bagging ration and etc. The DNN parameters include the number of hidden layers, the number of neurons, model training rate and etc. These parameters can be optimized using the advanced hyper-parameter optimizing algorithm such as Bayesian Optimization \cite{kong2018deep}. As our focus is not to develop new methods for parameter optimization, rather use the existing optimizer, the details of Bayesian Optimization are not included in this paper.  
\renewcommand{\algorithmicrequire}{ \textbf{Input:}}
\renewcommand{\algorithmicensure}{ \textbf{Output:}}
\begin{algorithm}[t!]
	\caption{fDNN chaining for the composition topology}
	\label{ag2}
	\scriptsize
	\begin{algorithmic}[1]
		\REQUIRE ~~\\
		$CT$: A composition topology with $M$ services\\
		$\mathbb{L^{S}}$: The set of all reputation indicators for a service $S$\\
		The training set $N^{S}$ containing $P$ samples for a service $S$;\\
		\ENSURE
		A chained network of fDNNs for the composition topology 
		
		\STATE Run DFS tree traversal on the $CT$ using the directed dependency graph.
		\STATE Sort the services in descending order of dependency depth
    	\FOR{$i = 1 \dots M$}
			\FOR{$j = 1 \dots M$}
				\IF{$i \neq j$}
				
				\IF{There exists a direct dependency between service $i$ and service $j$, i.e, service invocation from $i$ to $j$}
					\IF{There exists no fDNN containing either service $i$ or service $j$ as input layers}
					\STATE Create a new fDNN with input layer from both service $i$ and service $j$
					\ENDIF
					\IF{There exists a fDNN containing only service $i$ on the input layer}
					\STATE Create a new fDNN, $F^{new}$ with input layer from service $j$
					\STATE Find an existing fDNN $F^{i}$ that has input layer from service $i$ 
					\STATE Map the output of fDNN $F^{i}$ as an input for $F^{new}$
					\ENDIF
					\IF{There exists a fDNN containing only service $j$ on the input layer}
					\STATE Find existing fDNN $F^{j}$ that has input layer from service $j$ 
					\STATE Attach a new input layer to $F^{j}$ using service $i$
					\ENDIF  
				\ENDIF
			\ENDIF	
			\ENDFOR
			
		\ENDFOR
		\FOR{$i = 1 \dots M$}
		\STATE Create the DFR (unpruned random decision trees) with $\mathbb{L^{i}}$ and $N^{i}$ using Algorithm \ref{ag1}.
		\STATE Calculate out of bag (oob) error rate for each DFR.
		\STATE Select the first $k$ DFRs that has lowest oob error rate as $set(k)$
		\STATE Find the fDNN where service $i$ is one of the input.
		\STATE Map $set(k)$ on the input layer of fDNN for service $i$ 
		\ENDFOR
		\RETURN A chained network of fDNN according to composition topology
	\end{algorithmic}
    	%\vspace{-.2in}
\end{algorithm}

\subsection{Confidence of Bootstrapped Reputation}
The proposed fDNN training has uncertainties such as a fewer number of training samples, over-fitting or under-fitting. It is required to measure the confidence of a bootstrapped reputation which is not included in the training samples. The confidence value of a bootstrapped reputation specifies the reliability of the prediction. We identify two indicators that are important to measure the confidence of the bootstrapped reputation: 1) \textit{the overall accuracy of the fDNN} and 2) \textit{the prediction probabilities on the output layer of the fDNN for a given composition topology as input}. First, we calculate the overall accuracy of the fDNN, denoted as $ac$. If we have $a$ numbers of composition typologies with the actual reputation for testing purpose and the chained fDNN correctly predicts $b$ numbers of samples. The general accuracy $ac$ is computed using Equation \ref{Eqac}.

\begin{equation}
	\small
	\label{Eqac}
	ac = \frac{\text{The number of correctly predicted samples, } (b)}{\text{The total number of samples, } (a)}.
\end{equation}

We compute the bootstrapping probabilities on the output layer of the fDNN denoted as $bp$. Given an input sample to classify, the fDNN activates all the output neurons and the highest activation value is selected as the bootstrapped reputation class. The following example illustrates how the activation values of the output neurons may affect the confidence of the bootstrapped reputation. Let us assume, the output layer in a fDNN has only three reputation class, [`high', `moderate', `low']. There are two training input samples. A fDNN is trained with activation values, $A =[.4, .3, .3]$ for the first sample, and $B=[.9, .05, .05]$ for the second sample. Both $A$ and $B$ has the highest activation values on the `high' reputation and `high' should be selected as the bootstrapped reputation.  Although both cases produce the same result, the activation values are considerably different. In $A$, the conflicting neurons are almost activated which are attached to `low' and `moderate' classifications. However, in $B$, these neurons are not activated at all. Hence, the confidence of prediction using $B$ should be higher the confidence of prediction using $A$. We apply the \emph{softmax function} \cite{du2018towards} to extract the confidence information from the network. The softmax function is applied to the activation values of the $K$ output neurons and normalizes it into a probability distribution consisting of $K$ probabilities. If $x_i$ is the highest activation value of the $i^{th}$ neuron, the bootstrapping probabilities are computed as follows: 

\begin{equation}
\label{eq:bp}
    bp= \frac{e^{x_i}}{\sum_{i=1}^K e^{x_i}}, \; x_i \text{ is the highest activation value}.
\end{equation}

We consider the tuple $(ac, bp)$ as the confidence of the bootstrapped reputation. Note that $ac$ and $bp$ are correlated. The effectiveness of $bp$ which represents the uncertainty of reputation bootstrapping is influenced by the bootstrapping accuracy $a$. If $ac$ is quite low, the effectiveness of $bp$ is also low as it is unlikely that a non-trustworthy model will generate correct predictions for some random tests. In such cases, we only consider $a$ as an effective metric to evaluate the confidence of reputation bootstrapping. In addition, $bp$ can more comprehensively reflect the bootstrapping confidence when $ac$ has a higher value. 

\section{Experiments}

%We conduct a set of experiments to evaluate the proposed reputation bootstrapping approach.
We evaluate our proposed reputation bootstrapping approach by performing a set of well designed experiments. All the experiments are conducted on computers with Intel Core i5 CPU (2.13 GHz and 16GB RAM) and programmed in Python. We compare the proposed approach with the Topology Free Reputation Bootstrapping (TFRB), the Random Forest \cite{liaw2002classification} and the DNN approach \cite{silver2016mastering}. We consider the following three evaluation criteria:
\begin{itemize}
    \item \textbf{Accuracy:} It measures the prediction accuracy of different bootstrapping approaches for both atomic and composite services. We also analyse the confidence of the bootstrapped reputation.
    \item \textbf{Scalability:} It analyses how different approaches perform with increasing topology size and reputation classes. 
    \item \textbf{Runtime Efficiency:} It analyses the training time and prediction time of different approaches to bootstrap reputation. 
    
\end{itemize}

\subsection{Data and Parameter Description}
GitHub\footnote{\url{https://github.com}} is a widely used data set for a range of applications \cite{beller2017travistorrent}. Because of the modeling and behavior similarities to our research focus, we selected Github dataset as the appropriate platform to evaluate our proposed approach. We consider public repositories in Github as composite services. The \textit{insight} of a repository specifies both the dependent and dependencies or submodules which are publicly available. For example, the insight of the repository `mopidy/mopidy'\footnote{https://github.com/mopidy/mopidy} depicts that the repository has 237 watchers, 261 dependent repositories, 13 dependencies, 106 contributors, its community or owner has 22 other repositories with average 1000 ratings or stars and 120 followers. GitHub also provides insights about the commits, code frequency, and forks of the public repositories. The composite services are collected with a scrapper\footnote{\label{scrapper}\url{https://github.com/nelsonic/github-scraper}} as follows:
\begin{enumerate}
%[topsep=0pt,parsep=0pt,partopsep=0pt,leftmargin=10pt,labelwidth=6pt,labelsep=4pt]
\item Given a repository, we retrieve the dependent repositories from ``dependency insight'' page link. If this is the first repository in the topology and there are no dependencies, we discard the repository. Otherwise, we select at most 3 dependent repositories with uniform distribution.
\item For each of the dependent repositories, we perform step 1 to build the topology recursively. The depth of the recursion, i.e., the composition topology is set to 5.  As a result, each composition is formed with a maximum $(1+3+3^{6}+3^{9}+3^{12})$ repositories.    
\end{enumerate}

We collect 300 composite services which includes total 7555710 repositories and 10 reputation indicators for each repository. Hence, total number of data points is 75557100 which provides an opportunity to assess the performance in realistic settings. As the proposed approach uses a generic model to leverage the reputation-related factors, the diversity of reputation-related factors in the Github dataset is a perfect match to test the accuracy and scalability of the proposed approach. Experiments in different domains would provide a more robust evaluation of the proposed approach. However, finding the right data set is proved to be quite challenging as to the best of our knowledge, we could not find any public datasets other than Github that have reputation-related factors, component services' ratings, composition topologies and invocation history.

\textit{The ratings or stars of a composite repository is considered as the ground truth}. The corresponding class of the direct reputation is set by the $Lvl$ value using Equation \ref{Eq2}. The proposed approaches will be evaluated against this ground truth. We also collect the 10 reputation indicators for each repository using the GitHub scrapper\textsuperscript{\ref{scrapper}}. 
The GitHub API\footnote{\url{https://developer.github.com/v3/}} provides a keyword-based semantic search function to discover similar repositories. Repository descriptions are used to extract meaningful keywords. Similar repositories with the same keywords can be identified through the semantic search function provided by the Github API\footnote{\url{https://developer.github.com/v3/}}. The keywords are extracted from repository descriptions by removing stop words and duplicated words. 
The reputation indicators are grouped in 4 layers: a) \textit{provider layer (PL)}, b) \textit{community layer (CL)}, c) \textit{Similar Service layer (SL)}, and d) \textit{Insight layer (IL)}. Table \ref{tab:stat} reports the statistics of training data sets. We create 10 different sets with $Lvl = \{3, 5, 10, 15, 20, 25, 30, 35, 40, 45\}$ values. If a set contains $Lvl =20$, it states that the training set has 20 different classes. We conduct our initial experiments with $Lvl =20$ dataset set. We apply K-fold cross validation \cite{jung2018multiple} where $(k = 5)$, so there are 5 iterations of model training and testing. Each iteration contains different sets of 80\% data for training, and 20\% data for testing. All the methods are implemented in Python with packages Scikit-learn\footnote{\url{https://scikit-learn.org/stable/}} and Tensorflow\footnote{\url{https://www.tensorflow.org}}. Initially, the default parameters of these implementations are chosen in the experiments. Later, these values are tuned using a grid search with the training dataset. \textit{We compare the performance of the proposed fDNN with other classifiers under same or similar settings.}
    
\begin{table}[t!]
\vspace{-.2in}
\begin{minipage}{.45\textwidth}
\caption{Dataset Statistics}
  \vspace{-.1in}
  \label{tab:stat}
  \resizebox{1\textwidth}{!}{
    \begin{tabular}{ |p{5.3cm}|p{1.4cm}| } 
 \hline
 \textbf{Data type}&  \textbf{Statistics}\\ 
 \hline
 Number of total repositories  & 7555710 \\ 
 \hline
 Number of composite repositories  & 300 \\ 
 \hline
 Average component services/submodules  & 36  \\ 
 \hline
 Target: Average stars or ratings of a repository & 11.4 \\ 
  \hline
 PL: Average stars of an contributor & 12.4  \\ 
 \hline
 CL: Average stars of an owner &  16.4 \\ 
 \hline
 CL: Average stars of the other repositories of the owner &  12.7 \\ 
 \hline
 CL: Type of the owner & User, Organization  \\ 
 \hline 
  SL: Average stars of similar repositories &  7.8 \\ 
 \hline 
 SL: Average stars of the owners of similar repositories & 13.8 \\
 \hline 
 IL: Average number of watchers per repository & 35 \\ 
 \hline
 IL: Average number of dependents & 13  \\ 
 \hline
  IL: Average number of dependencies & 25  \\ 
 \hline
 IL: Average commits in a week & 1.4  \\ 
 \hline
 IL: Average forks & 5.1\\
 \hline
\end{tabular}
}
\end{minipage}
\hspace{.2cm}
\begin{minipage}{.51\textwidth}

\caption{Importance of reputation related indicators}
  \vspace{-.1in}
  \label{tab:stat1}
\resizebox{1\textwidth}{!}{
\begin{tabular}{ |p{4.25cm}|p{1cm}|p{1cm}|p{1cm}|} 
 \hline
 \textbf{Features}&  \textbf{MDA} & \textbf{MDCD} & \textbf{Avg.}\\ 
 \hline
 Provider layer & .181 & .111 &\textbf{.146} \\
 Average reputation of contributors & .181 & .111 &.146 \\ 
 \hline
 Community layer & .712 & .616 & \textbf{.664}\\
 Average reputation of an owner &  .194 & .166 & .18 \\ 
 Average reputation of other repositories of the owner &  .322 & .298 & .31\\ 
 Type of the owner - User  & .042 & .012 & .027\\ 
 Type of the owner - Organization & .153 & .098 & .125 \\ 
 \hline 
 Similar Service layer & .014 & .217 & \textbf{.115}\\
 Average reputation of similar repositories &  .003 & .007 & .005\\ 
 Average reputation of the owners of similar repositories & .011 & .021 & .016\\
 \hline 
 Insight Layer &.179& .201 & \textbf{.19}\\
 Average number of watchers & .057 & .061 & .059 \\
 Average number of dependents & .053 & .054 & .054 \\ 
 Average number of dependencies & .033 & .024 & .028\\ 
 Average commits in a week & .024 & .026 & .025\\ 
 Average forks & .012 & .036 & .024\\
 \hline
 
\end{tabular}
}    
\end{minipage}
\vspace{-.2in}
\end{table}

%\subsubsection{Model Parameters Tuning}

%\vspace{-.1in}
\subsection{Results}
\subsubsection{Importance of Reputation Indicators}
% \begin{table}[t!]
%   \caption{Importance of reputation related indicators}
%   \vspace{-.1in}
%   \label{tab:stat1}
% %\resizebox{.48\textwidth}{!}{
% \begin{tabular}{ |p{5.6cm}|p{1cm}|p{1cm}|p{1.7cm}|} 
%  \hline
%  \textbf{Features}&  \textbf{MDA} & \textbf{MDCD} & \textbf{Avg.}\\ 
%  \hline
%  Provider layer & .181 & .111 &\textbf{.146} \\
%  Average reputation of contributors & .181 & .111 &.146 \\ 
%  \hline
%  Community layer & .712 & .616 & \textbf{.664}\\
%  Average reputation of an owner &  .194 & .166 & .18 \\ 
%  Average reputation of other repositories of the owner &  .322 & .298 & .31\\ 
%  Type of the owner - User  & .042 & .012 & .027\\ 
%  Type of the owner - Organization & .153 & .098 & .125 \\ 
%  \hline 
%  Similar Service layer & .014 & .217 & \textbf{.115}\\
%  Average reputation of similar repositories &  .003 & .007 & .005\\ 
%  Average reputation of the owners of similar repositories & .011 & .021 & .016\\
%  \hline 
%  Insight Layer &.179& .201 & \textbf{.19}\\
%  Average number of watchers & .057 & .061 & .059 \\
%  Average number of dependents & .053 & .054 & .054 \\ 
%  Average number of dependencies & .033 & .024 & .028\\ 
%  Average commits in a week & .024 & .026 & .025\\ 
%  Average forks & .012 & .036 & .024\\
%  \hline
 
% \end{tabular}
% %}
% \end{table}
In the first experiment, we explore which indicators play an important role. Table \ref{tab:stat1} reports the averaged normalized importance of the factors on the GitHub dataset using Mean Decrease Accuracy (MDA) and Mean Decrease Confusion Degree (MDCD). The results demonstrate that every reputation-indicator influences the reputation bootstrapping with different degrees. The importance of the indicators is dependent on the application domain. The reputation-related indicator `community' has the highest importance to predict the reputation of service in GitHub. The experimental results also show that the factors similar service and component service have quite low importance in reputation prediction. The possible reasons may include: (1) semantic similarity cannot be applied to group repositories with similar reputations; and (2) the reputation of a composite repository is more influenced by its own developers rather than its sub module repositories. Hence, the `community' should be the higher layer in the framework. The `community layer' is formed with four indicators. In contrast, the other factors have relatively lower importance. There is only one indicator in the `provider layer'. Although the `insight layer' has greater importance than the `provider layer' (.19 to .146), the insight layer has five indicators. The similar service layer has two indicators and their individual importance is at the bottom in comparison with other layers. We also find that the average reputation of other repositories of an owner (.31) has greater influence than the average reputation of the contributors (.146). A possible explanation of this phenomena is that a contributor may get stars from other users based on his/her social relations (following or followed at (GitHub) which may not his/her reputation on a particular repository. Instead, the owner's past experiences on other repositories effectively reflect the owner's ability to providing a reputable repository.  
\begin{figure}[t!]
\vspace{-.2in}
   \begin{minipage}{.49\textwidth}
   \vspace{-.2in}
	%\centering
  	\includegraphics[width=1\textwidth,height=.9\textwidth]{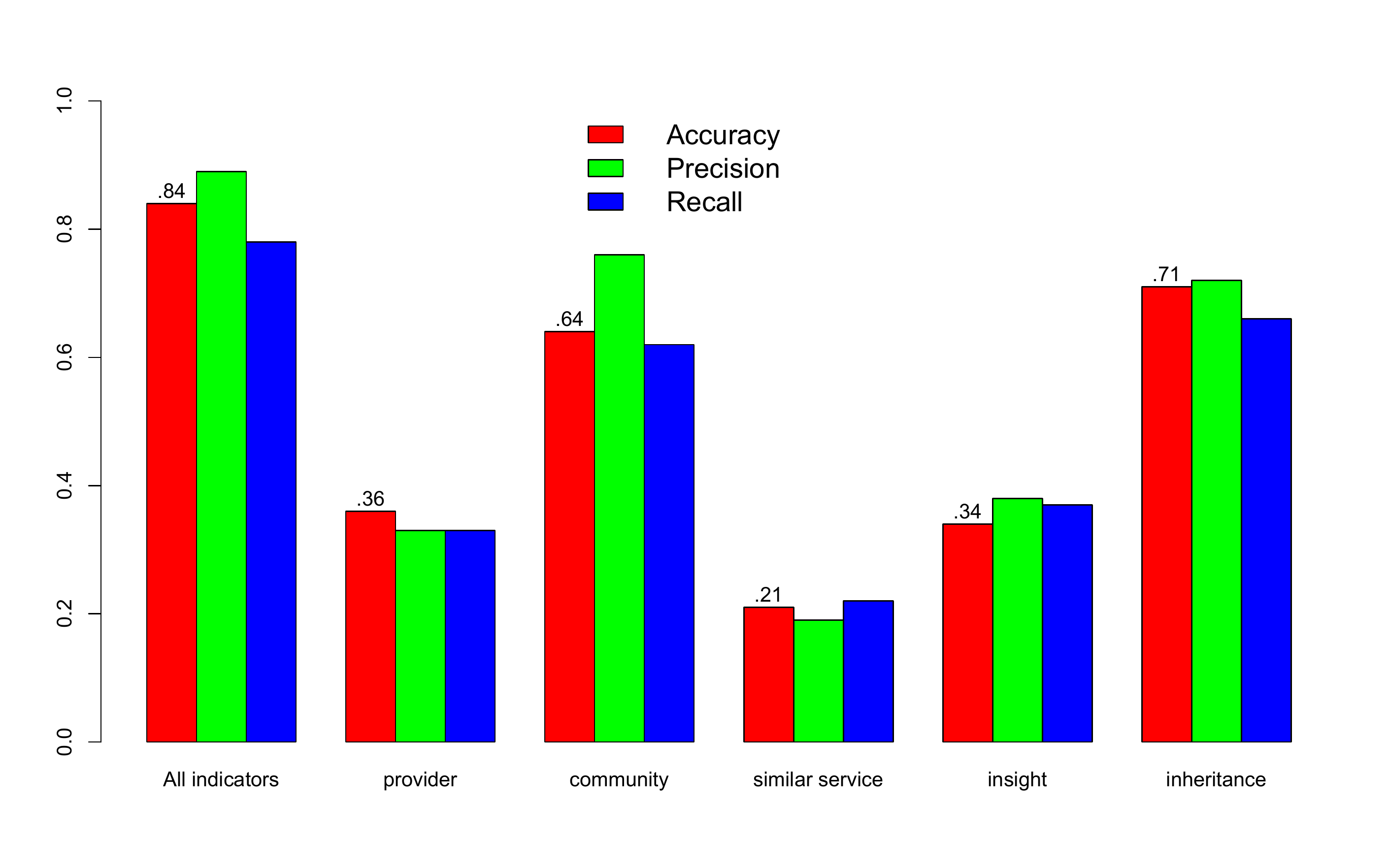} 
  	\vspace{-.25in}  
    \caption{Reputation bootstrapping accuracy for atomic services}
    \label{fig:Exp1}
   \end{minipage}
   \hspace{1mm}
   \begin{minipage}{.49\textwidth}
   \vspace{-.2in}
	\centering
  	\includegraphics[width=1\textwidth,height=.9\textwidth]{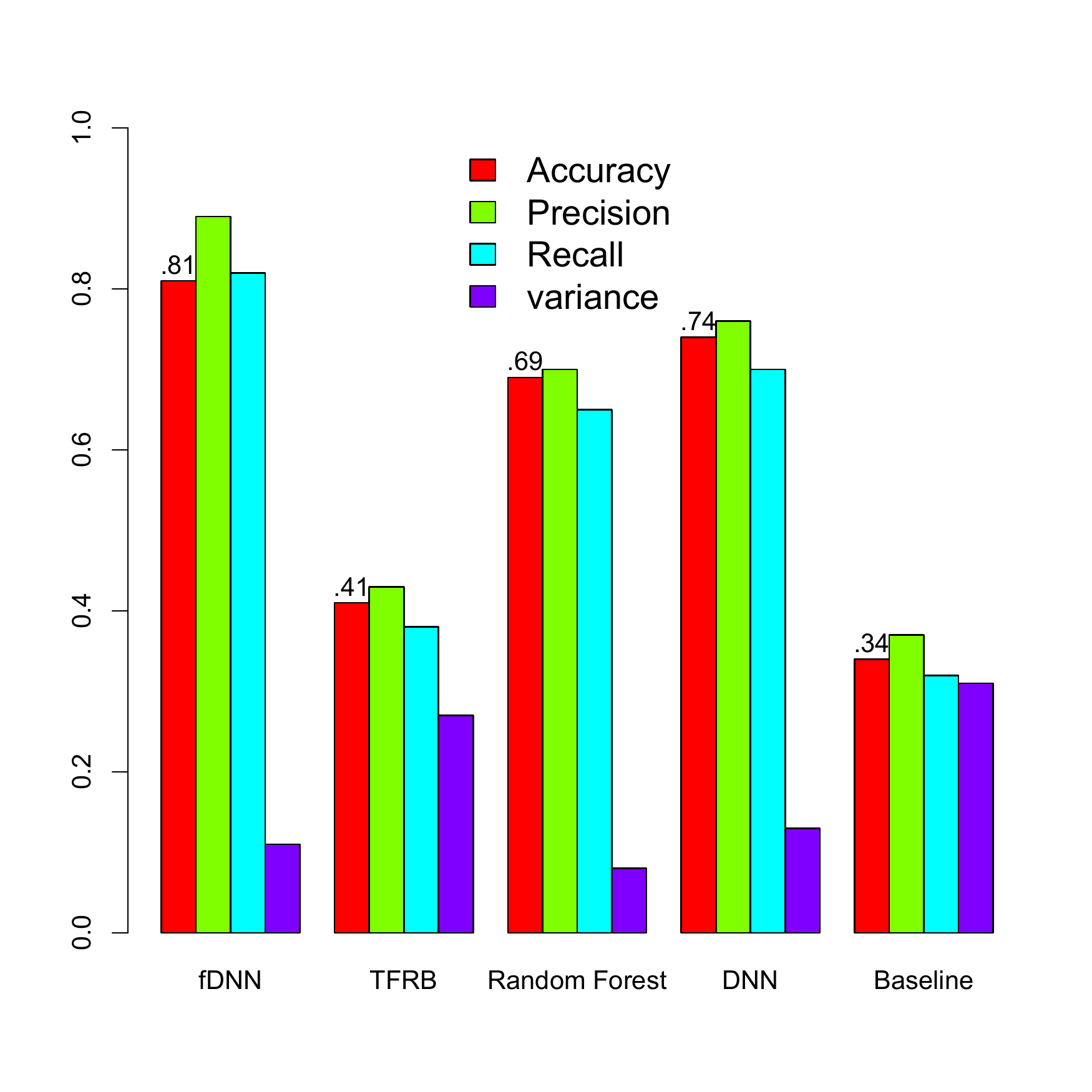} 
  	\vspace{-.25in}   
    \caption{Reputation bootstrapping accuracy for composite services}
  	%\vspace{-.1in}
	\label{fig:Exp2}
   \end{minipage}
   \vspace{-.2in}
\end{figure}

\subsubsection{Accuracy for Atomic Services}
Next, we evaluate the efficiency of proposed layer-based approach for atomic services. \textit{The stars or direct reputation of the repository is our ground truth}. We consider all reputation indicators together at first and then separately consider single layers, i.e., (`provider', `community', `similar service' and `insight') to train our model. We compare our approach with the inheritance mechanism \cite{nguyen2012bootstrapping}.  
% \begin{figure}[b]
% 	\vspace{-.2in}
% 	\centering
%   	\includegraphics[width=.65\textwidth]{figures/Exp1.pdf} 
%   	\vspace{-.2in}  
%     \caption{Reputation bootstrapping accuracy for atomic services}
%   %	\vspace{-.2in}
% 	\label{fig:Exp1}
% \end{figure}
% \begin{figure}[t!]
% 	\vspace{-.1in}
% 	\centering
%   	%\includegraphics[width=.5\textwidth, height=.33\textwidth]{figures/Exp2.pdf} 
%   	\includegraphics[width=.65\textwidth]{figures/Exp2.pdf} 
%   	\vspace{-.2in}   
%     \caption{Reputation bootstrapping accuracy for composite services}
%   	\vspace{-.1in}
% 	\label{fig:Exp2}
% \end{figure}
We consider three metrics to evaluate the efficiency: a) \textit{accuracy}: the ratio of correctly predicted observations to the total number of observations, b) \textit{precision}: the ratio of correctly predicted positive observations to the total number of positive observations, and c) \textit{recall}: the ratio of correctly predicted positive observations to the total number of observations. Figure \ref{fig:Exp1} shows that the accuracy, precision, and recall is high (avg. 8.5) when we consider all the layers. The dominant layer `community' has higher accuracy (approximates to .66) than the other layers. The single reputation indicators, i.e., the provider layer has .36 accuracy. The results reflect that the accuracy of a bootstrapped reputation is highly correlated with the importance of the reputation indicator. We also implement the inheritance mechanism \cite{nguyen2012bootstrapping}, which only considers the feature average reputation of the other repositories of the owner. The accuracy of the inheritance mechanism is close to the community layer, however, it is approximately 15\% less accurate than the proposed approach with all layers.

\subsubsection{Accuracy for Composite Services}
% \begin{figure}[b]
% 	\vspace{-.2in}
% 	\centering
%   	\includegraphics[width=.65\textwidth]{figures/Exp3.pdf}   
%     \vspace{-.1in}    
%     \caption{Effect of topology size on bootstrapping accuracy}
%   	%\vspace{-7mm}
% 	\label{fig:Exp3}
% \end{figure} 
In the third experiment, we compare the accuracy, precision and recall of the proposed topology-aware approach (chained FDNN) with the topology free approach (TFRB), the simple baseline, i.e., representing the minimum reputation of component services as the reputation of a composite service and two general classifiers, i.e., random forest and DNN. Note that both random forest and DNN follows that topology and are chained using the Algorithm \ref{ag2}. Figure \ref{fig:Exp2} shows that the accuracy of the proposed fDNN based approach is the highest (approximately .81). The accuracy decreases around 50\% to .41 for the TFRB. The random forest and the DNN have similar accuracy and close the fDNN (around 9\% less accurate). The simple baseline performs worst in the comparison with only 34\% accuracy of the reputation bootstrapping. The variance in the proposed model, random forest and DNN are also relatively lower (i.e., .11, .08, and .13 respectively) compared to the simple baseline and TFRB. The lower variance value in the 5-fold validation provides higher confidence on the accuracy of the proposed model. We conclude that the topology and important reputation indicators should be considered to bootstrap the reputation of a composite service.   
\begin{figure}[t!]
\vspace{-.2in}
   \begin{minipage}{.49\textwidth}
   \vspace{-.2in}
	\centering
  	\includegraphics[width=1\textwidth,height=.9\textwidth]{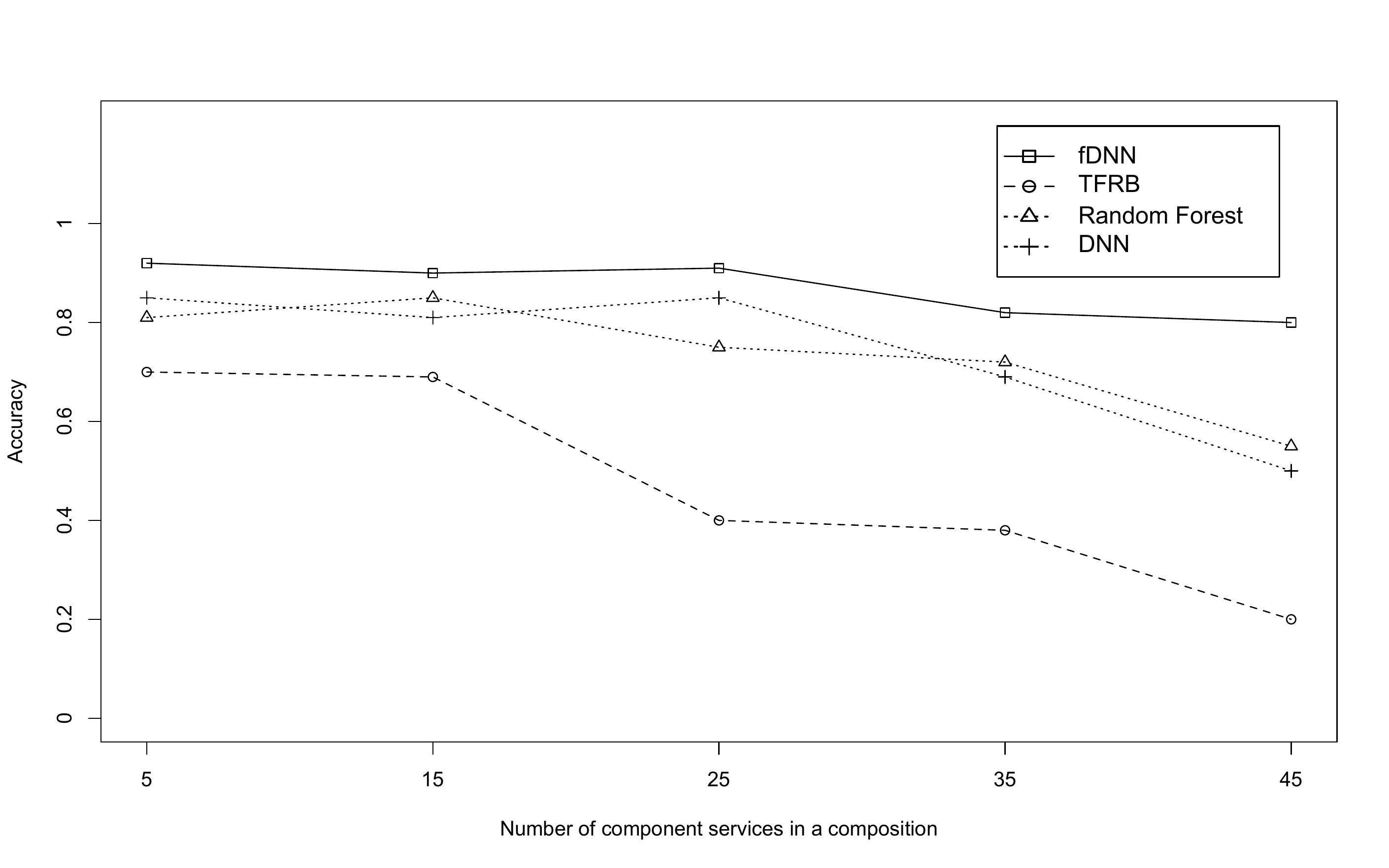}   
    \vspace{-.2in}    
    \caption{Effect of topology size}
  	%\vspace{-7mm}
	\label{fig:Exp3}
   \end{minipage}
   \begin{minipage}{.49\textwidth}
   \vspace{-.34in}
   \centering
  	\includegraphics[width=1\textwidth,height=.9\textwidth]{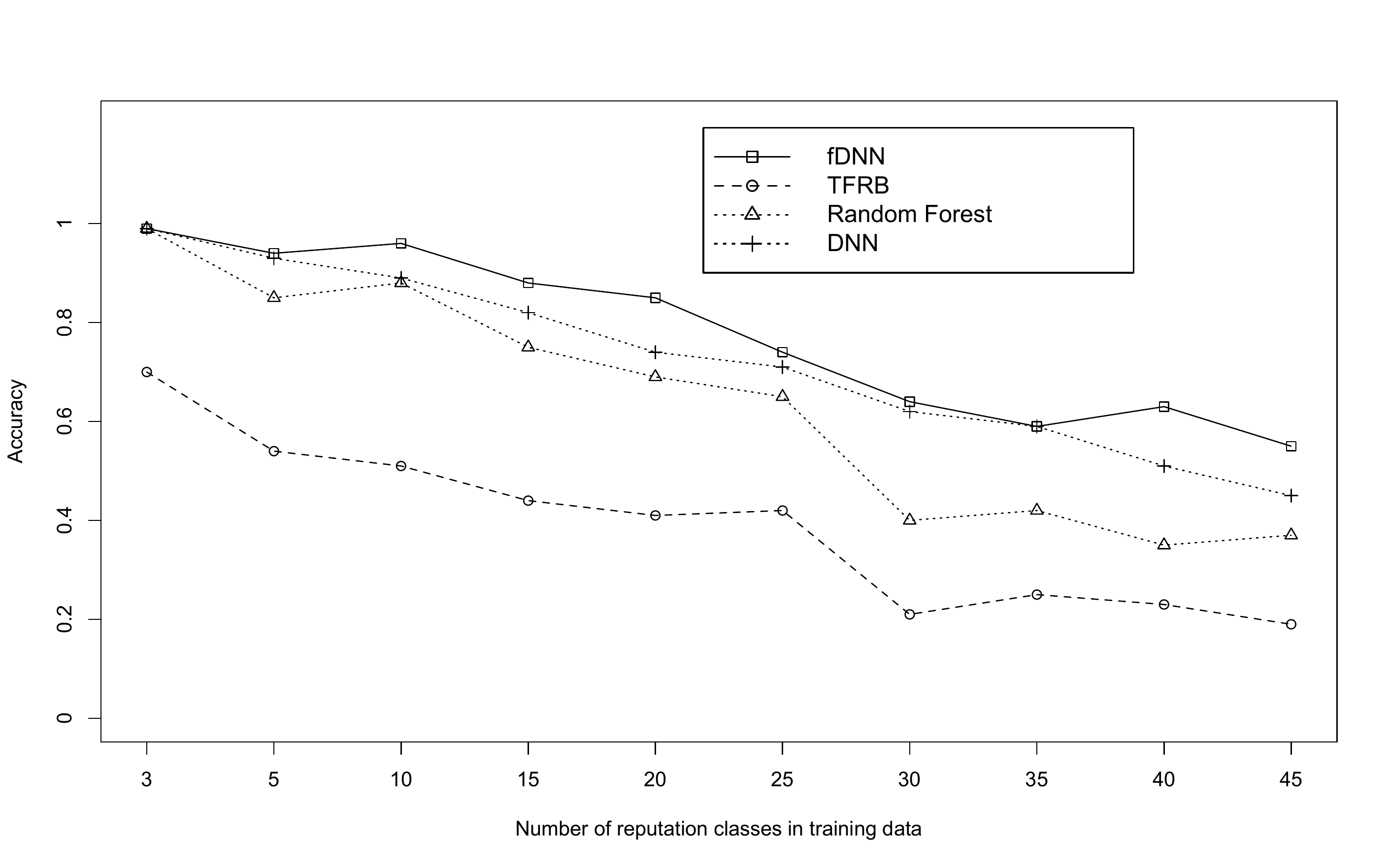}
  	\vspace{-.2in}   
    \caption{Effect of the number of reputation classes}
  	%\vspace{-.1in}
	\label{fig:Exp4}
   \end{minipage}
   \vspace{-.2in}
\end{figure}

\subsubsection{Effect of The Topology Size}
In the next experiment, we evaluate the effect of the topology size, i.e., number of component services in the reputation bootstrapping process. We consider five different size: (5, 15, 25, 35, 45). We create separate training sets for each size and they are trained separately. Figure \ref{fig:Exp3} demonstrates the scalability of the bootstrapping approaches, i.e., how the accuracy is affected when the topology size increases. Note that an increase in topology size also increases number of reputation-indicators in the training models without the increase of training samples. Hence, the prediction accuracy in smaller compositions may decrease as the models may not learn the increased interactions in the large compositions. Figure \ref{fig:Exp3} shows that all approaches have higher accuracy when the number of component services is very low. The reputation interactions among a small number of component services are not significant and we can bootstrap without considering the topology. The effect of reputation transfer is higher in larger compositions. As a result, the accuracy of TFRB decreases sharply when the composition gets bigger. Note that the decreasing rate of accuracy for the proposed fDNN is significantly lower than the other approaches. We conclude that the proposed approach is scalable. 

% \begin{figure}[t!]
% 	%\vspace{-5mm}
% 	\centering
%   	\includegraphics[width=.65\textwidth]{figures/Exp4.pdf}
%   	\vspace{-.1in}   
%     \caption{Effect of the number of reputation classes}
%   	\vspace{-.1in}
% 	\label{fig:Exp4}
% \end{figure}   

\subsubsection{Effect of The Number of Reputation Classes}
In this experiment, we evaluate the accuracy of the bootstrapping approaches on the 10 different training sets with different reputation classes. The reputation classes are used to transform the quantitative ratings (continuous) into qualitative values (discrete) for the model training. If the number of reputation classes increases, it represents a more accurate representation of the actual reputation value. A higher number of classes implies that the discrete qualitative reputation is semantically closer to the corresponding continuous quantitative value in practice. For example, when $Lvl = 3$, it semantically represents that there are only `high', `moderate' and `low' classification of the training samples. The lowest reputation class in a set of 3 reputation classes could be semantically interpreted to a broader range [0, .33]. In contrast, the value range of the lowest reputation class is significantly narrower, i.e., [0, .023] in a set of 45 reputation classes which provides a more accurate interpretation of the reputation. Hence, we consider up to $Lvl = 45$ for the reputation classification. Figure \ref{fig:Exp4} demonstrates the scalability of the bootstrapping approaches when the number of reputation class increases. Note that an increase in reputation class increases the output layer of the training models without the increase of training samples. The prediction accuracy with a smaller number of output reputation class (i.e., coarse-gained prediction) should be higher than the prediction accuracy with a large number of output reputation class (i.e., fine-grained prediction). Figure \ref{fig:Exp4} depicts that proposed fDNN approach results .8 or above accuracy up to $Lvl = 25$. However, its accuracy drastically decreases by $Lvl > 25$. In contrast, the accuracy of random forest and DNN drops from  .8 when  $Lvl > 15$. The accuracy of the TFRB drops from  .7 when  $Lvl > 3$. We conclude that the proposed fDNN can handle a larger number of reputation classifications.

\subsubsection{Confidence of The Bootstrapped Reputation}
The proposed bootstrapping approach predicts a reputation class with a confidence value $bp$ which is calculated using Equation \ref{eq:bp} in the testing phase. The predicted reputation may match with the actual observation (positive match) or may not match (negative match). Note that both the positive match and the negative match have confidence values. We have used 100 composite repositories in the testing where 73 samples have positive matches and 27 samples have negative matches. Figure \ref{fig:Exp5} shows the confidence distributions in positive and negative matches. We find that around 80\% positive matches have a higher confidence ranges (0.6 to 0.99). On the other hand, around 60\% negative matches have lower confidence ranges (0.01 to 0.59). %We conclude that a bootstrapped reputation with a higher confidence value has a higher probability to be accurate.
% \begin{figure}[b]
% 	\vspace{-.2in}
% 	\centering
%   	\includegraphics[width=.6\textwidth, height=.35\textwidth]{figures/Exp5.pdf}
%   		\vspace{-.2in}   
%     \caption{Distribution of confidence ranges}
%  % \vspace{-.1in}
% 	\label{fig:Exp5}
% \end{figure} 
\subsubsection{Runtime Efficiency}

\begin{figure}
\vspace{-.1in}
   \begin{minipage}{.49\textwidth}
   \vspace{-.2in}
	\centering
  	\includegraphics[width=1\textwidth]{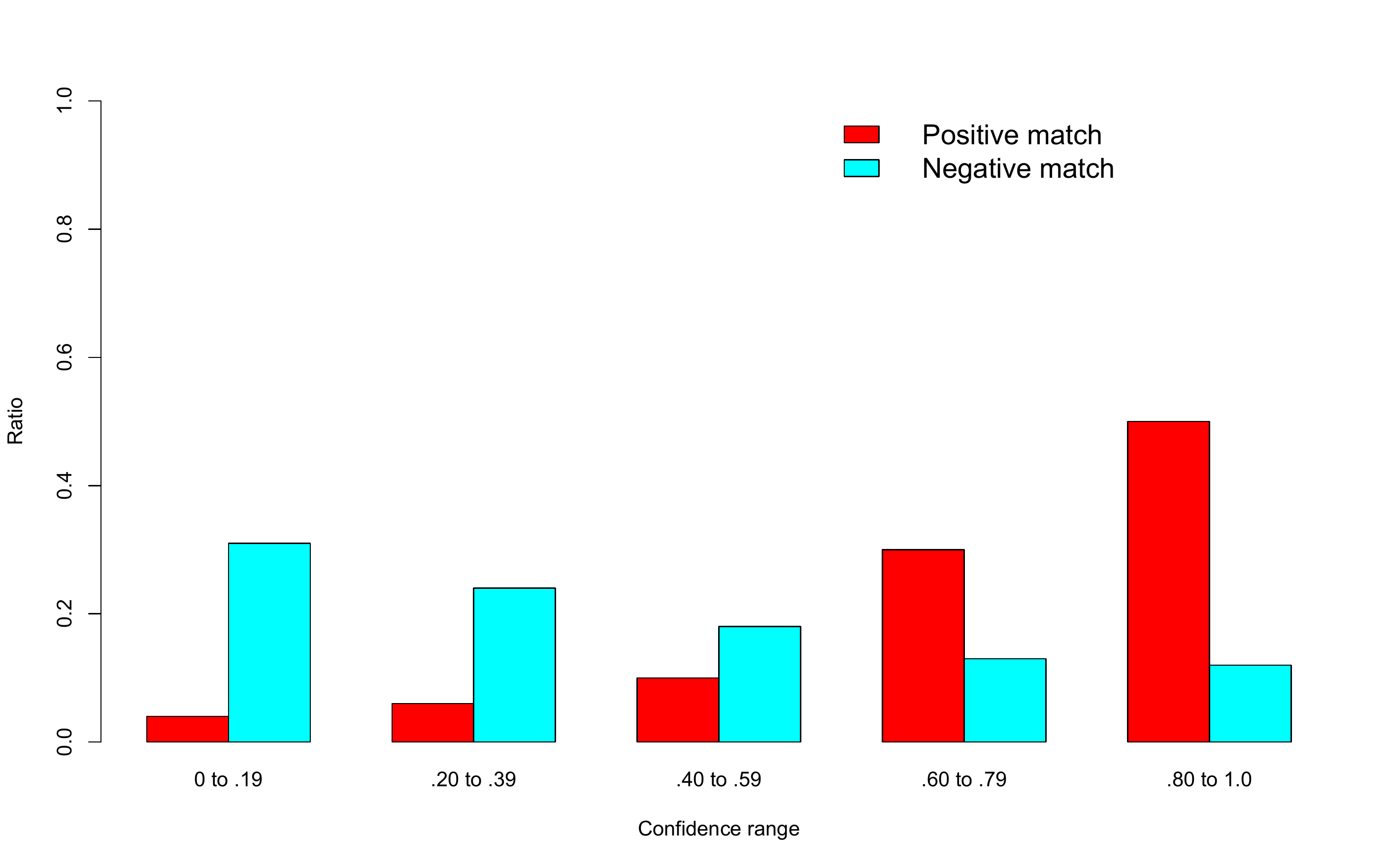}
  		\vspace{-.3in}   
    \caption{Distribution of confidence ranges}
 % \vspace{-.1in}
	\label{fig:Exp5}
   \end{minipage}
   \begin{minipage}{.49\textwidth}

  \vspace{1cm}
\resizebox{1\textwidth}{!}{
\begin{tabular}{ |c|c|c|c|c|} 
 \hline
&  \multicolumn{4}{|c|}{Time (hours)} \\ 
\hline
 \textbf{Method} &  \textbf{Set A} & \textbf{Set B} & \textbf{Set C} & \textbf{Average}\\ 
 \hline
 TFRB & 0.31 & 0.81& 2.12& 1.08\\
 \hline
 Random Forest & 0.82 & 1.63 & 3.61& 2.02 \\
 \hline
 DNN & 0.55& 1.89& 3.56 & 2.00\\
 \hline
 fDNN& 1.27& 2.26& 5.98& 3.17\\
 \hline
\end{tabular}
}
 \caption{Runtime Efficiency}
 \label{tab:stat2}
   \end{minipage}
   \vspace{-.2in}
\end{figure}

Figure \ref{tab:stat2} shows the model training time for three different sets, \textit{A}: 100 composite repositories (3461 component repositories), \textit{B}: 200 composite repositories (5966 component repositories) and \textit{C}: 300 composite repositories (8821 component repositories). The TFRB has lowest training time as it does not consider topologies. Both the DNN and the random forest has faster training time than the fDNN. The key reason is that fDNN has two phases of learning on random forest part and the DNN part. The training time increases if the training samples are increased for all the models except TFRB. We conclude that TFRB is the most time efficient.

% \begin{table}[htb]
%   \caption{Runtime Efficiency}
%   \vspace{-.1in}
%   \label{tab:stat2}
% %\resizebox{.48\textwidth}{!}{
% \begin{tabular}{ |c|c|c|c|c|} 
%  \hline
% &  \multicolumn{4}{|c|}{Time (hours)} \\ 
% \hline
%  \textbf{Method} &  \textbf{Set A} & \textbf{Set B} & \textbf{Set C} & \textbf{Average}\\ 
%  \hline
%  TFRB & 0.31 & 0.81& 2.12& 1.08\\
%  \hline
%  Random Forest & 0.82 & 1.63 & 3.61& 2.02 \\
%  \hline
%  DNN & 0.55& 1.89& 3.56 & 2.00\\
%  \hline
%  fDNN& 1.27& 2.26& 5.98& 3.17\\
%  \hline
% \end{tabular}
% %}
% \end{table}
           
\section{Related Work}
Computational trust and reputation models are widely researched topics in different domains, e.g., E-commerce, Recommendation systems, Communication Networking, Social Networking, E-health, Cloud Computing, IoT, Blockchain, and Service-oriented Computing (SOC) \cite{3236008}. Trust and reputation are important QoS to compose services in dynamic contexts \cite{3077584.3077590, 9284066, 8818422}. Our focus is the reputation bootstrapping of a composite service given its composition topology and the history of the reputation-related factors for the atomic or component services. To the best of our knowledge, existing approaches mainly focus on two directions: a) reputation bootstrapping for an atomic service, and) reputation propagation in a composite environment. The reputation bootstrapping approaches for an atomic service are typically classified into three categories: \textit{feature}-based, \textit{observation}-based approaches and methods to handle uncertainty.
\begin{itemize}
[topsep=0pt,parsep=0pt,partopsep=0pt,leftmargin=10pt,labelwidth=6pt,labelsep=4pt]
\item \textbf{Feature-based approaches:} The typical workflow of these approaches is to detect relevant reputation-related features, to build a bootstrapping model with the features using history, and to predict the reputation using the model for a new atomic service. A layer-based bootstrapping framework is proposed in \cite{qu2017confidence} that determines the importance of reputation-related indicators. Provider, community and similar services are considered as the indicators in the random forest-based learning forest. An artificial neural network (ANN) based model is proposed to predict the reputation of a new service by learning the correlations between reputation indicators and performance of existing services \cite{wu2015neural}. The reputation indicators are determined by the service features. The proposed approach in \cite{DBLP:conf/wise/SkopikSD09} identifies similar services as another reputation indicator using the crowdsourced tagging of different services. Social network analysis is considered as a key enabler to find similar services \cite{8108459}. Such analysis provides means to capture user behavior as well as historic features to bootstrap trust using the analytic hierarchy process (AHP). Both inheritance-based and referral-based approaches are designed to bootstrap the reputation in \cite{nguyen2012bootstrapping}. The inheritance mechanism uses the providers' existing performance as an indicator. The referral mechanism applies community referrals in the bootstrapping process. In \cite{DBLP:conf/icsoc/HuangLNFCT14} and \cite{DBLP:journals/dss/ZachariaMM00}, the reputation of a new service is derived using the mean and minimum trust value of all similar services from the similar provider (profile matched). Other important indicators of a new service's reputation are its service level agreement (SLA) and  quality of information (QoI) \cite{8455908}. If the SLA provides some forms of guarantee on the future performance, the initial reputation is assigned to a higher value \cite{DBLP:conf/atal/HuynhJS06,DBLP:journals/sigecom/MaximilienS02}. A continuous SLA monitoring method is proposed to dynamically adjust the reputation values of a cloud service in \cite{xiao2010reputation}. A new bootstrapping mechanism for evaluating the reputation of new web services is proposed based on their initial Quality of Service (QoS) attributes, and their similarity with existing web services that have long-term consumer feedback \cite{tibermacine2015regression}. The reputation-related indicators are fitted into a multi-dimensional regression analysis to predict the reputation of an atomic service. Note that supervised learning is not always possible if the trust data is sparse and unlabeled. The semi-supervised algorithm, i.e., Transductive Support Vector Machine (TSVM) algorithm is applied improved for sparse trust mining in \cite{3369390}.   

\item \textbf{Approaches to handle uncertainty}: The feedback from consumers may be collected from more than one source. For example, the review of an eBay product may be collected from eBay, Google product review or social networking sites. This increases the subjectivity of trust and creates uncertainty in reputation bootstrapping. A fuzzy multi-criteria decision-making model is proposed to identify the correlation of uncertain reputation-related factors and the critical success factors (CSFs) for the reputation value in \cite{8694843}. The Statistical Cloud-Assisted Reputation Estimation (sCARE) model is proposed in \cite{2792979} that incorporates the uncertainty and fuzziness of trust to bootstrap rater's credibility. The reputation is considered as the majority rating function considering all rater credibilities. the provenance of a rating score is considered as n important contextual meta-data in the computation of the ranks \cite{3183323}.          

\item \textbf{Observation-based approaches:} The typical framework of these approaches is to assign an initial reputation to a new service and then observe its behavior for a short period, i.e., trials or interact with the service to predict the fine-tuned reputation values. A bootstrapping approach is proposed using majority behavior in \cite{malik2009reputation}. The initial reputation assignment of an atomic service is evaluated by credible users (who use the service for the evaluation time) and their feedback is used to adapt initial reputation values. A reputation utility function is proposed to capture the behavior of a service and then it is matched with the initial reputation \cite{jiao2011framework}. An HMM model is proposed to capture the behavior sequences of the service and then it is matched against the initial assigned reputation value in the evaluation phase \cite{yahyaoui2011bootstrapping}. The stereotype of an atomic service is learned from the observations in a multi-agent environment and the reputation is calculated based on the expectation of future performance \cite{DBLP:conf/atal/BurnettNS10}.
\end{itemize}

There are three reasons why the approaches to bootstrapping the reputation for an atomic service could not be applied in compositions: a) they do not consider the composition topology, b) they do not consider the directional reputation influence among component services, and c) they do not consider the complex correlations of the reputation indicators among the component services. The approaches to bootstrapping the reputation for an atomic service only consider the reputation-indicators related to the service. If the directional reputation influence among component services, the composition topology, and the complex correlations of the reputation indicators among the component services are not considered, all the reputation indicators will be given the similar weight in bootstrapping process which may not represent the actual contribution level in the composition. 

%A reputation propagation approach is proposed to distribute users' ratings fairly among the component services \cite{nepal2009reputation}. In ad-hoc mobile networks, a graph-theory based approach is proposed to propagate reputation using agreement among mobile nodes \cite{liu2003reputation}. These approaches are not fit in our context as we require reputation propagation from component services to the composite service. 

%Our proposed approach is classified into the -based category. Compared to the other two categories, it requires no extra process (e.g., commitment management or a trial period). As a result, it is more practical and easier to achieve in real-world situations.

\section{Conclusion}
We propose a topology-aware reputation bootstrapping approach using reputation-related indicators for composite services. The random forest algorithm is applied to learn the layer-based importance of the reputation indicators for each component service. We propose the fDNN chaining approach to incorporate the reputation influence of the service invocation direction in a composition topology. The trained chained fDNN predicts the reputation of a new composite service by learning correlations among reputation indicators. We also evaluate the confidence of the predicted reputation. Experimental results with real-world dataset show that the proposed approach can effectively identify importance levels of reputation indicators. The proposed topology-aware approach is around 50\% more accurate the topology-aware approach. The chained fDNN for reputation bootstrapping is scalable and has acceptable runtime efficiency. Our proposed approach can be classified into the feature-based category. It is more practical in real-world situations as it requires no extra process of observations in a trial period. One of the key limitations of the proposed supervised approach is that the reputation indicators require to be labelled by experts before the model training. In future work, we will develop a data-driven bootstrapping approach using unsupervised learning where labels of reputation indicators are unknown. One of our assumption is that we have history or training data of different composite topologies. We will explore fDNN chaining without history or incomplete information in the future work. The proposed approach does not prescribe any composition strategy to use as it is outside the scope of this paper. We will also explore implementing other classification approaches, e.g., SVM, XGBoost, and LightGBM for reputation bootstrapping in the future work.

\begin{acks}
This research was partly made possible by DP160103595 and LE180100158 grants from the Australian Research Council, and ORS grant from Curtin University. The statements made herein are solely the responsibility of the authors.
\end{acks}

\bibliographystyle{ACM-Reference-Format}
\bibliography{reference.bib}

\end{document}